\documentclass{kapproc} 

\usepackage{procps}

\usepackage[dvips]{graphicx}

\nochapequationnumber 

 \let\footnote\savefootnote

\let\footnotetext\savefootnotetext

\let\footnoterule\savefootnoterule

\normallatexbib 

\let\lcitebracket[
\let\rcitebracket]

\begin{document}
\articletitle{Broken Symmetry and Coherence \\ of Molecular  Vibrations \\
                     in Tunnel Transitions}
\chaptitlerunninghead{Broken Symmetry and Coherence} \rhead{Broken Symmetry
and Coherence}
\author{Alexander M.Dykhne \altaffilmark{1,2}}
\altaffiltext{1}{TRINITI, 142092 Troitsk, Russia}
\author{Alexander G.Rudavets\altaffilmark{2}}
\altaffiltext{2}{Moscow Institute of Physics and Technology, 141700
Dolgoprudny, Russia} \email{Arudavets@mics.msu.su\footnote{address for
correspondences}}

\begin{abstract}
We examine the Breit-Wigner resonances that ensue from field
effects in molecular single electron transistors (SETs). The
adiabatic dynamics of a quantum dot elastically attached to
electrodes are treated in the Born-Oppenheimer approach. The
relation between thermal and shot noise induced by the
source-drain voltage $V_{bias}$ is found when the SET operates in
a regime tending to thermodynamic equilibrium far from resonance.
The equilibration of electron-phonon subsystems produces
broadening and doublet splitting of transparency resonances
helping to explain a negative differential resistance (NDR)of
current versus voltage (I-V) curves. Mismatch between the electron
and phonon temperatures brings out the bouncing-ball mode in the
crossover regime close to the internal vibrations mode. The
shuttle mechanism occurs at a threshold $V_{bias}$ of the order of
the Coulomb energy $U_c$. An accumulation of charge is followed by
the Coulomb blockade and broken symmetry of a single or double
well potential. The Landau bifurcation cures the shuttling
instability and the resonance levels of the quantum dot become
split because of molecular tunneling. We calculate the tunnel gaps
of conductivity and propose a tunneling optical trap (TOT) for
quantum dot isolation permitting coherent molecular tunneling by
virtue of Josephson oscillations in a charged Bose gas. We discuss
experimental conditions when the above theory can be tested.
\end{abstract}

\newpage
\begin{flushright}
{\sl This paper is dedicated to the\\ blessed memory of\\ A.P.Kazantsev}
\end{flushright}

\section{Introduction}

In this new millennium, we are witnessing the birth of molecular
electronics, which can now operate with a single atom or a
molecular nano-scale cluster called a quantum dot. Each quantum
state of the dot can be characterized by electron tunneling, since
the position of its discrete level is not averaged due to thermal
spreading $k_BT$, especially at cryogenic temperatures $T$ such
that $\Delta E\gg k_BT$, where the $\Delta E$ is the energy
spacing between levels and $k_B$ denotes the Boltzmann constant.
The quantum dot is confined between electrodes in a composite
nano-scale system. Typically, the molecular cluster is trapped on
the surface of a lead by a Lennard-Jones or van der Waals like
potential. The electronic reservoirs are set out of equilibrium by
applying a bias voltage $V_{bias}$, which modifies the
electrochemical potentials (Fermi levels) of each electrode $\mu
_\sigma $ driving a current through the quantum dot. The question
arises: What kind of the coherent properties one would expect by
measuring the electron transport in such system?

The experiments with nanoclusters, e.g. including fullerene
molecules $C_{60} $ \cite{Park} and $C_{140}$ \cite{Pasupathy},
which were addressed in a single electron transistor (SET), reveal
a plethora of new behavior, which cannot be explained within the
framework of solid-state nanostructures \cite{Datta} theory. For
instance, there are rich structures of conductivity resonance in a
magnetic field \cite{ParkJ} and anomalously high Kondo
temperatures, above $50K$, found in \cite{Yu}. The conductivity
gap is well correlated with the resonant frequency $\Omega =5$ mev
of the ''bouncing-ball'' mode of $C_{60}$ and manifests itself as
an enhancement of the conductance. There are two schools of
thought, which attribute this enhancement either to Frank-Condon
transitions or to the shuttling instability \cite{Gorel} -
\cite{Fedor} discussed below. For both schools the issue of
quantum mechanical coherence is the matter of a considerable
concern. We reconcile the above points of view in the adiabatic
approximation, i.e. by assuming that electronic degrees of freedom
are much faster than that of the dot. This is the case of local
equilibrium in phase space between the electronic flows and the
dot dynamics. Can the electron flow dephase the quantum dot by
scattering off it?

From first sight, there is a physical constraint for dephasing,
since an electron passing through the dot carries information in
one direction, which is fixed by the bias voltage. This produces
an asymmetry (chirality mapping) that accompanies the charging, by
virtue of the electron affinity to the quantum dot. The charging
modifies the Coulomb energy that competes with the elastic
deformation potential of the dot. Thence, the Coulomb potential
breaks the original symmetry of the molecular bonds in such a way
that the growth of $V_{bias}$ produces a bifurcation of the
bonding potential. The broken symmetry signifies the existence of
a border for the non-demolition quantum measurements. The
quantum-classical transition can thus be detected by the broken
phase of the oscillating current.

We have seen that the phase coherence of the oscillating current
is affected by the tunnel coupling to thermal reservoirs due to
the itinerant electrons. The phase-breaking process also changes
the interface states as a result of interaction with other
electrons via global Coulomb blockade. The Coulomb energy of
electrons and holes forces their coherent reorganization. The
electron coherence length $L_e$ increases with decreasing
temperature $k_BT$. As a result of freezing, $k_BT\le 1$ mK, a
solid-state system becomes mesoscopic on scales $L_e\le 1\mu m$.
\footnote{When the electron's phase coherence length $L_e$, i.e.
the typical distance for the electron to travel without losing its
phase, is of the order of the dot mean free path $L_d$, such a
system is called mesoscopic. The regime ensuring $L_e \gg L_d$ is
called ballistic.} Under this condition, the description of
transport in terms of the local conductivity breaks down. The
transport in a ballistic regime is dominated by electron
scattering from the interface but not from the quantum dots. Most
investigations of nanoscale systems are carried out for ballistic
dots, whose theory and experiment have together formed a new realm
of mesoscopic physics \cite{Datta}. For instance, observation of
the Kondo effect in single-atom and single-molecular junctions has
led to a promising field called spintronics \cite{Wolf}.

The SET device reported in \cite{Park} is an example of a double
tunnel junction system in which the quantum dot self-oscillates
between the leads. The mode softness  significantly influences the
electronic transport due to the effect of mechanical deformation
on the electrical properties. Nanoelectromechanics (NEM) ensures
that the electron's phase can be preserved over distances larger
than the dot size, thus giving rise to a quantum interference
which cannot be observed in macroscopic conductors. The
conductance of dots could inherit this quantum coherence, which
can manifest itself in superconducting current echoes \cite{Vion},
for instance. It means that a single molecular NEM-SET could, in
principle, display a high charge sensitivity, enabling
non-demolition measurements at the quantum threshold.

The experiments going under this title encompass a wide range of
the electro-mechanical devices from the macro- to nano-scale.
Coupling of a mechanical oscillator to  non-equilibrium baths is
accompanied by stiff dynamics of elastic self excitations and
brings the charge transfer into the shuttle regime as a result.
The evident advantage of shuttling lays in avoiding the tunnel
coupling bottleneck. The phenomenon of the shuttle instability for
quantum dots in double junctions resembles a more general class of
the adiabatic quantum pump for electroacoustic and/or photovoltaic
effects \cite{Gloria}.

The current control at the one-by-one electron accuracy level is
feasible in mesoscopic devices due to quantum interference. Though
the electric charge is quantized in units of $e$, the current is
not quantized, but behaves as a continuous fluid according to the
jellium electron model of metals. The prediction of the current
quantization dates back to 1983 when D. Thouless \cite{Thoul}
found a direct current induced by slowly-traveling periodic
potential in a 1D gas model of non-interacting electrons. The
adiabatic current is the charge pumped per period, $I$, and has to
be multiple of the electron charge, i.e. $I=e\nu N$, where the
frequency $\nu =V/A$ is related to the traveling wave velocity $V$
and the wavelength $A$. Then the charge transmitted in the
adiabatic pump is period-independent.

The frequency dependence of the current $I$ holds true for a dot
shuttling between the leads and thereby modulating in phase the
conductivity of the sequential tunnel junctions. The electron
interference is manifested in an instability of the dot charge,
which is subject to either stochastic or periodic oscillations.
These shuttle dynamics allow one to find a new compromise between
tunnel charging and Coulomb forces, rigidity and elasticity of dot
bonds. Yet, a simple scattering theory that would help to estimate
a measurements of current induced by the shuttling is absent, to
our knowledge. Even for the usual I-V curve, characterized by
regions of negative differential resistance and observed
experimentally in \cite{Park}, a mutual consensus between the
Franck - Condon picture of electron transport and the shuttling
mechanism has not been found.

In the quest for a new functionality of NEM SET, the shuttling
mechanism has attracted a considerable interest as an effective
method of control of electron transport, whose current depends on
the frequency $\nu $. A description of the shuttling instability
can be based on a general master equation \cite{Gorel} and Green's
function \cite{Fedor} methods of coarse grained dynamics over a
scattering spectrum, without paying special attention to resonance
field effects. Almost all of the obtained results are strongly
model dependent and do not shed light on the underlying physics.
For example, remaining unexplained are the higher Fano factors in
the shuttling regime as compared to that obtained in the tunneling
regime\cite{Nord}. This is one reason why it is worthwhile to
develop a conceptually clear picture of the phenomenon, in the
spirit of the Breit-Wigner theory of resonance cross sections.
Another reason for this is that many results obtained both in the
framework of the scattering approach and by classical methods in
mesoscopic physics are applicable on an equal footing to the
shuttling process\cite{Blant}.

The universality of both the Breit-Wigner method \cite{Blant}
\cite{Kadig} and the shuttling is manifested by studies of the NEM
Josephson junctions \cite{Gorel}, \cite{Isac}. The latter belong
to the mesoscopic system wherein the Cooper-pair box is shuttled
between remote electrodes in the superconducting SET (SSET). The
NEMs favor coherent coupling\cite{Isac} and allow the suppression
of quantum fluctuations of dissipationless persistent current in
the ground state of the system. The shuttle mechanism reduces the
Fano factor at low temperatures of about $1\mathop{}mK$
\cite{Gorel}. This is in accordance with the general rule
\cite{Blant} that the voltage, the current, and the charge
oscillations due to Josephson plasmons are less noisy and more
entangled for strongly correlated system.

The shuttle mechanism based on the tunnel Hamiltonian\cite{Gorel}-
\cite{Fedor} is equivalent to the simplest possible Holstein-type
polaron models \footnote{the coupling of Einstein phonons to
finite electron density in conduction band.}. These models ignore
all complexity of the real molecular SETs: A detailed
understanding of the charge screening, geometry of electrodes,
hybridization with continuous and bound surface states, scattering
off impurities is either absent or presented in a fragmented
manner. The lack of knowledge about frequency shifts of the
scattering resonances is filled by a phenomenological approach.
Taking for granted their argumentation, we tackle the tunnel
resonances using a common theory of resonance scattering in the
Breit-Wigner approximation with a more pragmatic goal. By
developing the Born-Oppenheimer adiabatic strategy in forbidden
(for electrons) inter-electrode zone, where quantum dots are
allowed to move classically, we present here the current and the
quantum-dot charge in a self-consistent, model independent, and
tractable form.

In Sec. 2, we discuss the Landauer formula for the current,
deriving it in parallel with inferring the mean charge from
detailed balance conditions. This is followed by deducing the
resonant tunneling in the Breit-Wigner approximation, which
provides a convenient framework for description of quantum
transport phenomena. In Sec. 3, the dissipative tunneling is taken
into account with the help of a phenomenological model. In Sec. 4
we make use of adiabatic dot dynamics in order to describe field
splitting and broadening of the resonance levels. The main aim is
to show how do the dot oscillations between the leads influence
markedly the current-voltage curves. With an increasing voltage
the charging regimes change. Section 5 is dedicated to the Coulomb
blockade modified by the adiabatic motion in phase space. We
calculate the self-consistent charge accumulated in resonance
windows of conductivity in order to illustrate the tunnel term
transformation from a single well symmetry to a double well
symmetry. We analyze how the shuttling instability depends on bias
voltage. In Sec. 6 we estimate the shot noise of the shuttling
adopting the resonance scattering approach. In Sec. 7 we justify
the assumption of broken symmetry of the tunnel terms by employing
a quantum treatment of the SET setup in the Born-Oppenheimer
adiabatic approach. In Sec. 8 we propose the use of TOT protection
of a quantum dot from parasite hybridization with the substrate
surface in order to avoid dissipative tunneling roadblocks.
Coherence of electron transport via double wells is sketched
briefly in Sec. 9. Conclusions are given in Sec. 10.

\section{The Landauer formula.}

The Landauer's seminal suggestion, that the current is transmission,
dominates in mesoscopic physics and has applications to a variety of
systems, including the electron transport in solids, liquids, quantum wires
and dots. First and foremost, this theory describes
 conductance by purely dissipationless electrons scattering.
Pursuing the NEM phenomena, we shall follow a similar reasoning,
omitting the non-elastic effects on the microscopic scales.

Consider a ballistic quantum dot between two metallic terminals.
Their electronic reservoirs are held at the thermal equilibrium
described by the Fermi-Dirac distribution
\begin{equation}
f_\sigma (E)=\frac 1{{1+e^{(E-\mu _\sigma )/k_BT}}},
\end{equation}
where the chemical potentials $\mu _{l(r)}=E_F\pm eV/2$ correspond
to a shifted Fermi energy, $E_F$, while $eV$ is the biased
electron potential across the source and drain leads. Electrons
flow from the high potential $\mu _l$ to the low potential $\mu
_r$ passing the Fermi level $E_F$. The electronic scattering state
$\psi $ with the Fermi energy $E$ is normalized to unit flux in
the lead $\sigma \in (l,r)$ at far asymptotic distances $\chi =\pm
\infty $ from the quantum dot placed at a fixed coordinate $x$.
The partial current of electrons, averaged over the momenta
$p(E)=\sqrt{2mE}$, where $m$ is electron mass, is defined as
\begin{equation}
I_\sigma (\chi ,t)=\frac em{\mathop{\rm
Im}\nolimits}\int\limits_0^\infty {\frac{{dp(E)}}{{2\pi }}\mathop
{\psi _\sigma }\limits^{*}(\chi ,t)\frac{{d\psi _\sigma (\chi
,t)}}{{d\chi }}}f_\sigma (E)\mathop\cdot
\end{equation}
The measured current from the source to the drain electrode is
proportional to the energy integral of squared modulus of the
scattering matrix $\left| {S(E)}\right| ^2$, whose integrand is
called the quantum-dot transparency
\[
\Upsilon (E)=\left| {S(E)}\right| ^2,
\]
overlapped with the difference of Fermi functions of the electrons in the
right and the left leads
\begin{equation}  \label{Lan}
I=I_r(\chi =\infty ,t)+I_l(\chi =-\infty ,t)=\frac{{\pi e}}\hbar
\int\limits_0^\infty {dE\Upsilon (E)}(f_r(E)-f_l(E)).
\end{equation}
Here $\hbar $ is Plank's constant. The only quantum property
playing a role in the average current $I$ is the Pauli principle,
which dictates that each quantum state in the Fermi sea has to be
occupied by a single electron. This means that only a fixed number
of electrons can be accumulated in the scattering sector at a
fixed energy $E$, thus setting a limit on the average current flow
$I$. Factor $2$ in Eq. \ref{Lan}, representing the electron spin
degeneracy, has to be carefully taken into account, especially for
the detailed balance conditions, as demonstrated in Sec. 3 and
used in Sec. 5. At small bias, that is at $V\ll k_BT$, one writes
 $I=GV$, by introducing the linear conductivity $G$ defined as
\begin{equation}  \label{Con}
G=\frac{{2e^2}}h\int\limits_0^\infty {dE\left| {S(E)}\right|
^2}\frac{{df(E)} }{{dE}}\mathop\cdot
\end{equation}
The Fermi-Dirac distribution, $f(E)$, is the step-like function of
energy $E$, while its derivatives are delta functions in energy.
One therefore obtains
\[
G=g_0\left| {S(E_F)}\right| ^2.
\]
The conductance quantum $g_0=\frac{2e^2}h$ is a universal factor
of maximum conductivity for a single scattering channel at unit
transmission. The minimum quantum resistance is the inverse
quantity $g_0^{-1}=12.9K\Omega $, which implies that the
dissipation of an ideally transparent quantum dot occurs due to
the scattering at the interface with electronic thermal
reservoirs. This means that even for the system in a perfectly
ballistic condition, the coupling of the electron with the
reservoir induces decoherence. Irreversibility of an open system
arises from uncorrelated ''itinerant'' electrons broadening the
resonant state. ''Itineracy'' destroys the unitarity of the
quantum mechanical evolution, and the scattering matrix $S$ in the
Landauer theory is reduced to phenomenological determination.
Below, we present the Breit-Wigner resonance approximation for the
scattering matrix as an example of such an approach.

\subsubsection{Resonant tunneling in Breit-Wigner approximation}

Let us consider the tunnel resistance of a $\chi
=1\mathop{}nm$-broad vacuum gap between two gold electrodes. For
the Fermi energy $E_F\sim 8eV$ and the work function $W\sim 5eV$,
the dominating electron wave function exponentially decreases in
vacuum with the rate $\gamma =\sqrt{2Wm}/\hbar \sim
1.3\mathring{A}^{-1}$, where $m$ is the electron mass. The
enormous resistance $R_{1nm}=g_0^{-1}e^{-2\chi \gamma }\sim
2.5\mathop {}10^{15}$ Ohm prohibits  electron transport in vacuum
in the absence of a quantum dot.

For a dot fixed between two electrodes, a weak electron coupling
results from the tunnel current passing through the dot and
broadens the resonant dot levels. The broadening width
exponentially decreases as function of the distance $\chi $
between the dot and the electrode surface. The typical tunnel
probability for an electron to jump from the electrode to the
quantum dot formed by a $C_{60}$ molecule \cite{Park} placed at
the distance $\chi \approx 6\mathring{A}$ from the lead, can be
estimated as $\Gamma _0\sim E_Fe^{-2\gamma \chi }\sim 0.1-1\mu
eV$. In the situation when the quantum dot can move, its
hybridization with the lead depends on the dot coordinate
$\chi=\pm x$
\begin{equation}  \label{Gam}
\Gamma (\chi )=\Gamma _0e^{-2\gamma x}.
\end{equation}
The dot is characterized by the two parameters $\Gamma ^r=\Gamma
_0e^{2\gamma x}$ and $\Gamma ^l=\Gamma _0e^{-2\gamma x}$
describing the exponentially decreasing couplings $\Gamma ^{l,r}$
with the reservoirs $l$ or $r$. The total broadening
$\Gamma=(\Gamma ^r+\Gamma ^l)/2$ denotes the width of the energy
level $E_{dot}$. The broadening $\Gamma $ may vary with the level
energy and depends on the electron-electron correlation, although
not strongly. In the weak-coupling limit, the resonance width
$\Gamma $ is much smaller than the average spacing between the
energy levels $\Delta $, $ \Delta \gg \Gamma $. In this limit only
the scattering energies $E$ which are nearest neighbours to the
dot energy level $E_{dot}$ contribute to conductance. In this
case, the Breit-Wigner resonance of the scattering matrix
$|\mathrm{S}(E)|^2$ or the transparency $\Upsilon (E)$ reads
\begin{equation}  \label{BW}
\Upsilon (E)=|\mathrm{S}(E)|^2=\frac{{\Gamma ^r\Gamma ^l}}{{
(E-E_{dot})^2+\Gamma ^2}}\mathop\cdot
\end{equation}
The scattering matrix elements implicitly depend on $x$ via the
parametric coordinate dependence of $E_{dot}$ and $\Gamma $. The
tunnel coupling $\Gamma $ homogeneously broadens the resonance
energy levels $E_{dot}$. Therefore, the Lorentzian distribution of
the scattering matrix element implies an exponential decay via the
coupling of electrons to the environment across only a resonance
window of density of states $\rho (E)$
\begin{equation}  \label{Lor}
{\rho }(E)=\frac 1\pi \frac \Gamma {{(E-E_{dot})^2+\Gamma ^2}}\mathop\cdot
\end{equation}
By substituting the matrix $\mathrm{S}(E)$ from \ref{BW} into
Landauer's formula \ref{Lan} at zero-temperature limit, the
tunneling conductance is exactly represented as
\begin{equation}
\mathrm{G}_{\mathrm{L}}(E_{\mathrm{F}})=g_0\frac{{\Gamma ^r\Gamma
^l}}{{(E_{\mathrm{F}}-E_{dot})^2+\Gamma ^2}}\mathop\cdot
\end{equation}
For the most of the experiments of interest, the temperature $kT_B$
typically exceeds the resonance width $\Gamma \ll kT_B\ll \Delta $, and
hence the interference effect of adjacent levels separated by the $\Delta $
is inessential. The conductance $G$ from \ref{Con} is given by the integral
of the Breit-Wigner transparency \ref{BW} yielding
\begin{equation}  \label{Cond}
\mathrm{G(E}_{\mathrm{dot}}\mathrm{,T)}\approx
\frac{{\mathrm{G}_0}}{{ \mathrm{cosh}^{\mathrm{2}}\left( \Delta
_T\right) }},\mathop {} \nolimits_{}\Delta
_T={\frac{{\mathrm{E}_{\mathrm{dot}}\mathrm{-E}_{\mathrm{F
}}}}{\mathrm{2k_BT}}},\mathop
{}\nolimits_{}\mathrm{G}_0=\frac{{\pi g_0}}{{4k_BT}}\frac{{\Gamma
^r\Gamma ^l}}\Gamma ,
\end{equation}
where $\Delta _T$ is offset from the resonance energy from the
Fermi level in units of $k_BT$. The relationships \ref{Cond}
between the conductance $G$ and the tunneling broadening $\Gamma $
solve the electron transport problem for the static dot in the
ohmic regime.

The concept of the static dot is justified by the fact that the
dot's translational and rotational degrees of freedom vary slowly
as compared to the fast motion of the electrons. The parameters
$E_{dot}$ and $\Gamma $ of the scattering matrix $\mathrm{S}(E)$
represent adiabatic variables, validated by the Born-Oppenheimer
approach. An instant conductivity of a movable dot is to be
computed in a phase space of the NEM oscillator's coordinates $x$
and momenta $p$. The $x,p$ point plays the role of a partial
scattering channel in which the Wigner delay time of electron
tunneling $\tau =\frac{{\partial S(E)}}{{\partial E}}$ is the
shortest time scale. The marginal delay $\tau $ justifies the
picture of an instant scattering, where the current is obtained by
averaging over the phase space. This reasoning will thread
throughout this chapter after a short discussion of the proximity
effects and electron balancing flows in Sec. 3.

\subsubsection{The proximity effects and dissipative tunneling}

If the dot sticks to a metal lead and forms an ''adatom'', its
energy levels $E_{dot}$ get broadened by hybridization with the
surface continuum states. Since the electron affinity is different
for the dot and the host surface, the charge transfer causes a
change in the electrostatic potential inside the dot and shifts
the energy levels by a contact potential. The result is that the
adatom could be directly charged classically, by the electrons
flowing in and out of the conduction bands. For example, Rubidium
($Rb$) is electropositive on a gold ($Au$) surface, since its
ionization potential $ I_i\sim 4.2eV$ is smaller than the $Au$
work function $W\sim 5eV$. But it remains neutral on an isolator,
such as glass, because the bonding originates from the van der
Waals covalent coupling. The dot's charging depends on conditions
of ''quasi-equilibrium'' in the open driven system ''dot $+$
leads''.

For molecular dots placed far from the surfaces, the tunnel
coupling is rather weak, as compared to the mechanical and
electrostatic energies. Under these circumstances, the transport
is usually subject to global Coulomb blockade. When the charge
attempts to tunnel, it strongly perturbs the surface states,
inducing a coherent reorganization of electrons and holes
generating  phonons or plasmons on the substrates and
leads\cite{Braig}. Figure \ref{edm} shows how the Fermi-edge is
disturbed due to the tunneling.

A priori, when the phonon relaxation is faster than the tunneling
rates, thermodynamic equilibrium should hold at the temperature of
the host reservoir. However, for the nano-junctions the local
surface temperature may differ from the bulk equilibrium
temperature. This is due to the Anderson orthogonality catastrophe
(AOC)\footnote{ The notion of the OC \cite{AOC} was introduced by
P.W. Anderson with respect to the Fermi-edge singularity where the
overlap of states which differ in the number of holes by one tends
to zero in the thermodynamical limit.} associated with interplay
between the van der Waals and the electrostatic forces. The
electron tunneling affects the overlap between differently shifted
phonon ground states of the surface. The faster the tunneling
rate, the closer is the phononic overlap to zero, and that hinders
relaxation of the surface temperature. AOC presents the mechanism
also affecting
 the thermal state of the electronic reservoir due to
electron-phonon coupling. In Sec. 4, from comparison of our
theoretical I-V curves at different electron-phonon temperatures
and the experimental data \cite{Park} we infer that AOC exists.

In Sec. 3, we make use of the detailed balance conditions for
derivation of a generic expression for the mean charge of the
quantum dot. We formulate the field effect on the splitting and
broadening of the tunnel resonance for adiabatic evolution. Here
we want to emphasize that the irreversible decay is especially
important for establishing steady states in the non-equilibrium
system of the dot at contact with an external infinite bath.
Besides the aforementioned AOC, damping of the shuttle motion of
charged particles between metallic electrodes can be related to
radiative decay mechanisms, discussed long ago in the seminal
paper \cite{KS} by Kazantsev and Surdutovich.

\section{Current at detailed balance}

The system made-up of a quantum dot and two leads of different
Fermi energies experiences electron flows tending to bring the
system to thermodynamic equilibrium. In the steady-state regime,
the net current summed over the electrodes is zero, that is the
incoming and the out-going flows of electrons through the quantum
dot compensate each other. The balancing process is provided by
the tunnel rates $\Gamma ^\sigma $ weighted by the Heisenberg time
needed for an electron of energy $E$ to escape into the electrode
$\sigma =\in (r,l)$,
\begin{equation}
\Gamma ^\sigma =\frac \hbar {\tau ^\sigma }\mathop\cdot
\end{equation}
If we denote the electron distribution function inside the quantum
dot by $f_{dot}(E)$, the out-flow current at energy $E$ to the
right electrode is
\begin{equation}
I_{out}(E)=\frac{f_{dot}(E)}{\tau ^r}=\frac{\Gamma ^rf_{dot}(E)}\hbar \cdot
\end{equation}
The electrons populate the level $E$ of the quantum dot by
backward transitions from the lead with an incoming flow
\begin{equation}  \label{curin}
I_{in}(E)=\frac 2{\tau ^r}\rho (E)f_r(E),
\end{equation}
occurring through the resonance window $\rho (E)$ over the time
$\tau ^r$ , while the factor $2$ in the Eq. \ref{curin} accounts
for the spin degeneracy in the Fermi sea. The difference of the
in- and out-going electron flows at energy $E$ produce the net
current
\begin{equation}
I_r(E)=e\frac{\Gamma ^r}\hbar \left[ 2\rho (E)f_r(E)-f_{dot}(E)\right] .
\end{equation}
Analogously, the expression for the current on the left lead reads
\begin{equation}
I_l(E)=e\frac{\Gamma ^l}\hbar \left[ 2{\rho }(E)f_l(E)-f_{dot}(E)\right]
\cdot
\end{equation}
Since the detailed balance principle implies that in equilibrium the
electron population of the dot has to be at a steady-state, $I_l+I_r=0$, we
immediately arrive at the relation
\begin{equation}  \label{detbal}
\Gamma ^r\left[ 2\rho (E)f_r(E)-f_{dot}(E)\right] =\Gamma ^l\left[
f_{dot}(E)-2\rho (E)f_l(E)\right] ,
\end{equation}
which yields the distribution $f_{dot}(E)$
\begin{equation}  \label{chaden}
f_{dot}(E)=2\rho (E)\frac{\Gamma ^rf_r(E)+\Gamma ^lf_l(E)}{\Gamma ^r+\Gamma
^l}\mathop \cdot
\end{equation}
The net current in the scattering channel $E$ is represented as
\begin{eqnarray}
I(E)&=\frac{e\Gamma^l }{\hbar}(2\rho(E) f_l - f_{dot}(E)(E))\mathletter{a} \\
&=\frac{e \Gamma^r\Gamma^l}{\hbar \Gamma}\rho(E)(f_l(E)-f_r(E))\mathletter{b}
\\
&=\frac{2e}{h}\Upsilon(E) (f_l(E)-f_r(E)). \mathletter{c} \label{steadstat}
\end{eqnarray}
We see that the detailed balance is equivalent to  Landauer's
formula Eq. \ref{Lan}. The generic method of the steady state
regime is utilized here in order to emphasize a common meaning of
Landauer's ansatz of Eq. \ref{Lan} corresponding to Eq.
\ref{steadstat}. The current flows and electronic distribution
$f_{dot}(E)$ in the quantum dot are self-consistently related
provided
\begin{equation}  \label{char}
Q=2e\int\limits_0^\infty {dE}\rho (E)\frac{\Gamma ^rf_r(E)+\Gamma
^lf_l(E)}{\Gamma ^r+\Gamma ^l}\mathop\cdot
\end{equation}
The total charge accumulated in the resonance window is a
trade-off between the Fermi seas of the electrodes, allowing for
the establishment of equilibrium between the continuous spectra of
their conduction bands and the intermediate resonance state of the
quantum dot. The steady-state kinetics refers to diverse transport
problems, and it can be derived by more subtle techniques such as
the density matrix approach in the framework of the Keldysh
representation \cite{Keldysh} on a closed time path
\cite{Gogolin}, or the nonlinear Green function methods
\cite{Wingreen}, \cite{Datta}. But our aim is more pragmatic: We
apply Eq. \ref{char} in the Breit-Wigner approximation for
adiabatic motion of the quantum dot and calculate the average
current.

\subsubsection{Field splitting and broadening of the resonance level}

Due to the huge difference of the molecular and the electron mass,
slow molecular vibrations can be considered as quasistatic when
compared to the fast electron motion. The slowness of the
molecular vibrations justifies the Born-Oppenheimer adiabatic
approach. This strategy provides us with a paradigm useful for
consideration of electron tunneling through a movable quantum dot.
By the analogy to the Born-Oppenheimer molecular terms, we use the
concept of tunnel curves, representing a total electronic energy
as a function of the dot coordinate in the inter-electrode
regions, forbidden for the electrons. However, the massive dot
could move classically in this region, thus transporting an
attached electron and having a potential energy, called the tunnel
curve. Note that this tunnel term, being very similar to molecular
one, includes both the potential and kinetic electron energies as
a function of the dot center-of-mass coordinate $x$.

Having this in mind, the tunnel term can simply be regarded on
phenomenological grounds. From the other hand, the
quantum-mechanical method of Sec. 7 exemplifies an ''ab-initio''
approach employed here for the calculating the tunneling current.
The tunnel term's topology defines the adiabatic dot dynamics and
the average position of the measured electron. Thus, the averaging
of the instant scattering S-matrix or the resonance transparency
$\Upsilon (E(t))$ over the adiabatic paths becomes a basic
ingredient of the electron transport problem.

Without loss of generality, we set the resonance energy level
$\varepsilon _0 $ equal to the Fermi energy, $E_F=\varepsilon _0$.
In equilibrium, the dot assists the tunneling in remaining
neutral, since the electro-chemical potentials of the leads and
the dot are identical. The self-consistent ''electro-chemical''
potential for charging is $U_{mean}=U_c(Q-Q_0)=0$, where $U_c$ is
the Coulomb energy cost per one electron, and $(Q-Q_0)$ is an
extra charge with respect to unbiased electronic reservoirs. To
estimate the field effect we write down the mechanical and
electrostatic energy $E_{dot}$ as follows
\begin{eqnarray}
& E_{dot} = \varepsilon_0+ U _{mean} + E_{vib} +U_{ext}\mathletter{a}
\label{Edot} \\
&E_{vib} = p^2/2M + M \Omega^2 x^2/2 \mathletter{b} \\
& U_{ext}(x) = -F x \mathletter{c},
\end{eqnarray}
where $E_{vib}$ is the vibrational energy of the dot of mass $M$. The
''bouncing-ball'' mode consists of the kinetic part $\frac{p^2}{2M}$, and
the potential part $\frac M2\Omega ^2x^2$, with $\Omega $ being the
vibration frequency. The external field force is $F=eV/D$, where $D$ is the
tunnel (inter-electrodes) gap, and $V$ is the bias voltage.

If discharging and relaxation are faster than charging, the
quantum dot remains neutral. The main role of the intermediate
quantum dot is to assist a virtual tunneling through its resonance
state. The quantum dot in a simple harmonic model just oscillates
over a closed trajectory having the coordinate and momentum
\[
x_n(t)=x_{n0}\cos (\Omega t),\quad p_n=p_{n0}sin(\Omega t),
\]
where the $x_{n0}=x_0\sqrt{2n}$ and $p_{n0}=p_0\sqrt{2n}$ are integrals of
motion of $n$-th vibrational energy eigenstate. The zero-point amplitude and
momentum
\[
x_0=\sqrt{\frac \hbar {M\Omega }},\quad p_0=\frac \hbar {x_0}
\]
correspond to the vibrational ground state. By denoting the frequency shift
due to external potential field as
\[
\nu _n(t)=\frac{Fx_n(t)}\hbar ,
\]
we count off the dot energy levels $E_{dot}(t)=\hbar (n\Omega -\nu
_n(t))$, from the resonance reference point $\varepsilon _0=E_F$.
The detuning frequencies $n\Omega -\nu _n$ define the energy
positions of the transparency resonances, required for calculation
of the instant electron current by the Landauer's formula Eq.
\ref{Lan}. A thermodynamic average of the electron transport over
the dot vibrations has to be taken in order to obtain the mean
current and to compare the calculated I-V curves with experimental
current-voltage characteristics.

At increasing voltage, the coupling between the dot and the leads
tears the phonon temperature away from the equilibrium temperature
$T$. Thermalization of this driven open system involves
complicated dynamics of the electron-phonon interactions. As an
estimate, we use the Callen-Welton theorem in order to represent
the fluctuation-dissipation relation between the noise power and
the effective temperature of the system ''leads$+$dot''. On the
one hand, the Johnson noise\footnote{the fluctuations at
thermodynamic equilibrium are related in a universal way to the
kinetic response according to the fluctuation-dissipation theorem
(FTD)} reads
\begin{equation}  \label{sn}
S_{Jonh}=4k_BTG,
\end{equation}
where the $G$ in the right hand side of Eq. \ref{sn} is the
conductivity \cite{Blant}. The parameter $T$ has the meaning of an
effective temperature $T_{eff}$ for the non-equilibrium system. On
the other hand, the voltage $V$ produces the Schottky's noise that
gives the main contribution to the power spectrum at a low
temperature, except of the transparency resonances \cite{Blant}
\begin{equation}  \label{noise}
S_{Schottky}=2eI,
\end{equation}
where Ohm's law reads
\[
I=VG.
\]
By equating the power spectra of the thermal noise and the non-equilibrium
shot noise
\[
S_{Jonh}=S_{Schottky},
\]
we immediately relate the effective temperature $T_{eff}$ with the
voltage $V $, which drives the current and heats the SET to the
temperature
\begin{equation}  \label{Teff}
k_BT_{eff}=\frac{eV}2.
\end{equation}
This fundamental relationship has been obtained from the density
matrix of the quantum point contact coupled with mechanical
oscillator in \cite{Mozyrsky}. Therein quantum heating and damping
of the oscillator manifest themselves in an induced
quantum-classical transition. In fact, our simple FDT arguments
also indicate that heating by the electron-phonon interplay can be
much larger than heating by the thermal reservoirs. The
source-drain voltage $V$ adversely affects  the current and may
result in undesirable artifacts, such as electro-migration, SET
disruption, etc. Therefore, a non-demolished experiment requires a
small voltage and low temperature, where the adiabatic
approximation holds. In this limit, we can ignore the dependence
of tunneling probability $\Gamma $ both on the energy $E$ and the
coordinate $x$, setting $\Gamma \sim \Gamma _0$ near $x=0$, since
$\gamma x_0\ll 1$. For the coherent model, the instant value of
$\Upsilon (E,t)$ is given by
\begin{equation}
\Upsilon (E,t)=\frac 1Z\sum\limits_{n=1}^\infty {\frac{{\Gamma
_0^2e^{-\frac{{\hbar n\Omega }}{{k_BT}}}}}{{(E-E_d(n,t))^2+\Gamma
_0^2}},}
\end{equation}
where $Z=\sum\limits_{n=1}^\infty {e^{-\frac{\hbar n\Omega }{k_BT}}}$ is the
vibrational partition function. The path averaging yields the mean
transparency
\begin{eqnarray}
& \overline {\Upsilon (E)} = \frac{\Omega }{{2\pi }}\oint
{dt\Upsilon (E,t) }
\mathletter{a} \\
&= \frac{{\Gamma _0 }}{Z}{\mathop{\rm Im}\nolimits} \sum\limits_{n
= 1}^\infty {{csgn(c) e^{ - \frac{{\hbar n \Omega}}{{k_B T}}} }}
{{[(E - \hbar n \Omega - i\Gamma _0 )^2 - 2n(\hbar\nu_0)^2
]^{-\frac{1}{2}}}}\mathletter{b},  \label{transb}
\end{eqnarray}
where $csgn(c)$ \cite{Abramowitz} denotes a signum function of
complex argument $c=e^{-i\phi }$, and $\phi =arg(E-n\hbar \Omega
+\sqrt{2n}\hbar \nu _0-i\Gamma _0)$ is used to keep continuity of
the square root of the complex-valued function. The root function
originates from the uniform distribution of the particles along
the adiabatic path in the phase space $x,p$. The main contribution
to the resonances gives the phase space sector $p=0$, since the
electron has a higher probability to scatter off the ''static''
dot, than at the dot moving in-between the turning points. The
shifted frequencies $\nu _n$ are maximum at the turning points,
that provides the doublet splitting of the vibrational resonances
$n\Omega $. Since the frequency shift depends on the $x_n$, the
spectral broadening increases with the vibrational energy
$n\Omega$. The inset in the figure \ref{fig1} shows the resonance
transparency \ref{transb} averaged over the dot path.

In order to make comparison with the experiment \cite{Park}, we
chose the frequency $\Omega =5\mathop{}$ $meV$, which corresponds
to a $C_{60}$ quantum dot interacting with gold electrodes via the
Lennard-Jones potentials. For the electron temperatures
$k_BT=0.4meV$, the vibrational amplitude is $x_0 \sim 0.03
\mathring{A}$, and hence only the zero-point mode is active,
provided $\Omega \gg k_BT$ . The equilibrium distance $D/2\sim 6.2
\mathring{A}$ separates the center-of-mass of the quantum dot from
both leads. When $eV\le \Omega $, the frequency shifts $\nu _n$
are negligible, since the system ''leads$+$dot'' is cold $k_BT\ll
\Omega $, and therefore $\nu _n=Fx_n=\Omega x_n/D\ll n\Omega $.

However, on average, the current through the vibrating dot differs
significantly from the ''static'' dot transmission. At growing effective
temperature \ref{Teff}, the bouncing-ball mode produces \textit{an
inhomogeneous} broadening. This field effect results from averaging over the
frequency shifts $\nu _n(t)$ that depend on the location of the quantum dot
bouncing between the electrodes via the source-drain voltage characteristic,
while the tunnel coupling broadens the resonance line \textit{homogeneously}.

The Breit-Wigner approximation allows one to represent the
resonance tunneling spectrum as superposition of Lorentzian lines.
In the limit of narrow electronic levels $\Gamma _0\to 0$, the
resonance transparency obeys the equation
\begin{eqnarray}
&\overline {\Upsilon (E)} = \frac{{\Omega
\Gamma_0}}{{2Z}}\sum\limits_{n = 1}^\infty {\oint {dt} \delta (E -
E_{dot} (n,t))e^{ - \frac{{\hbar n\Omega}}{
{k_B T}}} }\mathletter{a} \\
&= \frac{{\Omega \Gamma _0 }}{{2Z}}\sum\limits_{n = 1}^\infty {e^{
- \frac{{\hbar n \Omega}}{{k_B T}}} \left. {\frac{{dt}}{{dE}}}
\right|_{E=E_{dot} (n,t)}} \mathletter{b}.  \label{tp}
\end{eqnarray}
This equation states that the resonance transparency of trapped dots, atoms
or molecules, is formed in the neighborhood of the turning points, where
$\dot E=0$. At these points, $p=0$, and the resonance transparency $\Upsilon
(E)$ becomes infinite in the limit $\Gamma _0\to 0$. However, the coupling
with the environment produces a finite broadening $\Gamma $ that negates the
divergence of the transparency $\Upsilon (E)$.

\section{Current Voltage Curves}

The adiabatic treatment of the dot's motion can be used to clarify
as yet unexplained features in the experimental I-V curves
\cite{Park}. To this end one not only needs to find the adiabatic
paths but also to perform averaging over their thermal
perturbations. In our first example, the electron reservoirs are
strongly coupled with the dot vibrations that are characterized by
the effective temperature $k_BT_{eff}=eV/2$. The quantum dot can
locally heat the surface of the leads up to the same temperature
$T=T_{eff}$. The reason behind the anomalous electron heating is
the electron tunneling perturbing the surface. The local
quasi-particle perturbations, e.g. plasmons, cannot relax fast
because of the AOC. The surface plasmons create no overlap between
the shifted phonon states. As a result of phonon-plasmon interplay
\cite{Braig} the electron temperature $T$ equilibrates with the
phonon temperature $T_{eff}$.
\begin{figure}[ht]
{\caption{The I-V curve at the same temperatures of electron and
phonon reservoirs due to the AOC effect. The inset shows the
doublet spectra caused by vibrating the dot with $\Omega=5meV$ and
$T_{eff}$ at the end point of the I-V curve.} \label{fig1}}
\sidebyside {\includegraphics[width=2.4in]{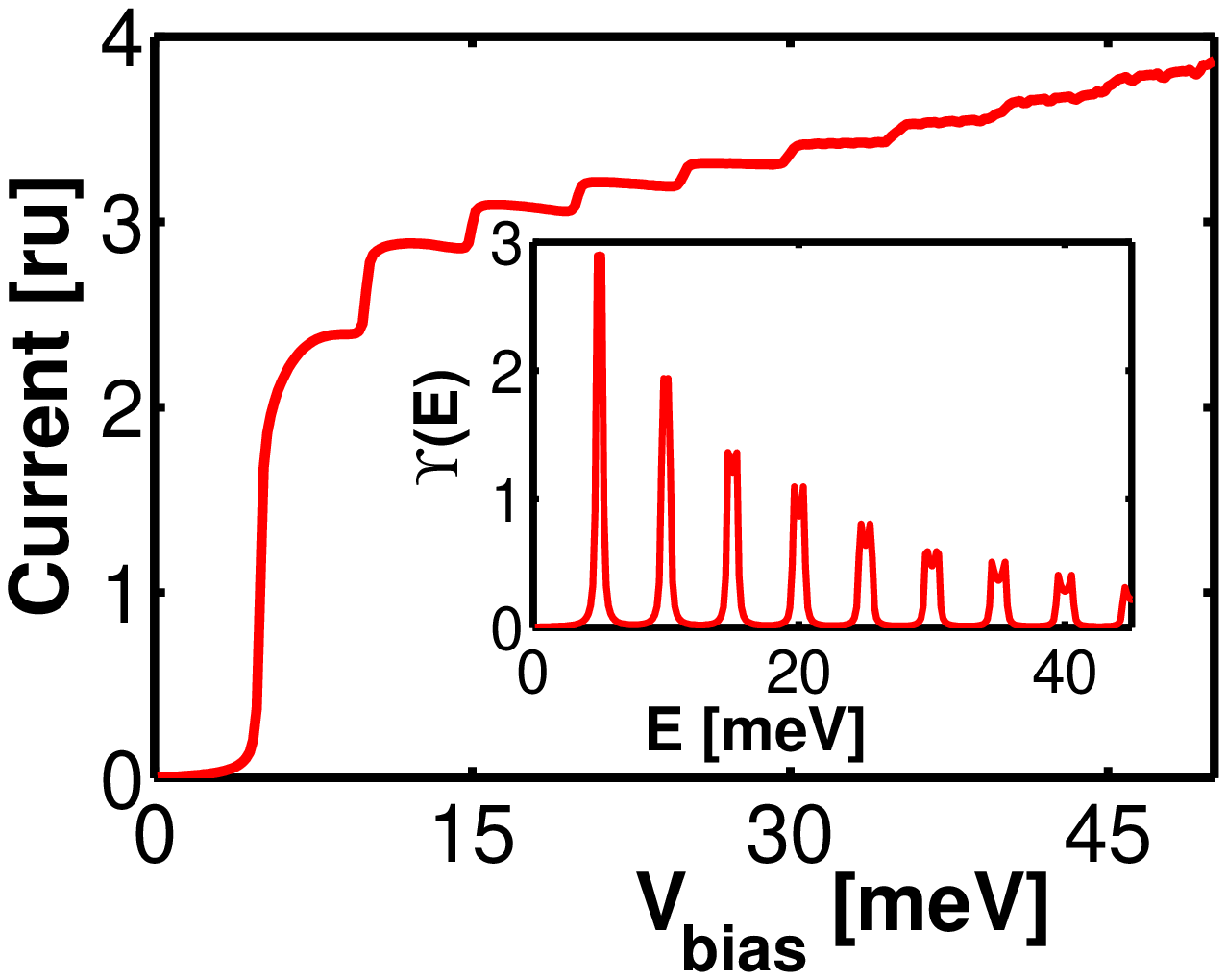}}
{\includegraphics[width=2.4in]{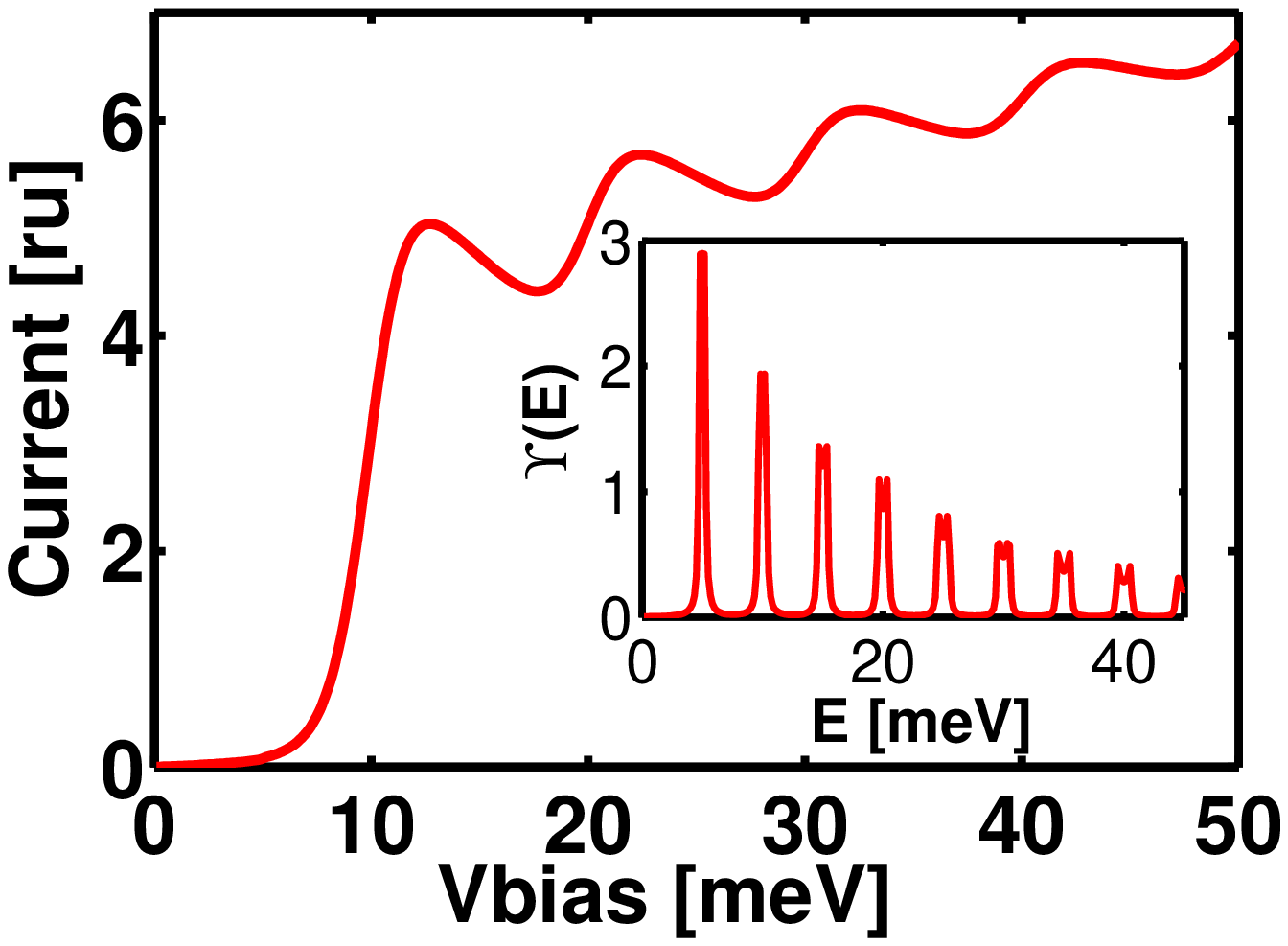}}
 \sidebyside {} {} {\caption{The I-V curve of  "dot$+$leads" system at the
equilibrium electron temperature $k_B T_e=0.4$ meV, but at a different
effective phonon temperature $k_B T_{ph}=eV/2$.}\label{fig2}}
\end{figure}

The electronic current versus the source-drain voltage shown in the figure
\ref{fig1} displays the typical features of $C_{60}$ SET \cite{Park} at the
open gate $V_{gate}\ll \Omega $. At a small gate voltage, surface charge
\cite{Braig} is allowed, and the AOC mechanism may take place. For the
$C_{60}$ SET, the phonons manifest themselves in the step-plateau-like
current, with the conductivity gap corresponding to the phonon frequency
$\Omega=5meV$, and with a negative differential resistance (NDR) regions.

The NDR can be explained by the field effect, which splits the
resonance scattering lines because of the frequency shifts at the
turning points, and therefore the electron flow  only partly
contributes to the current. In accordance with the Pauli
principle, the Fermi energy spectrum window given by the
difference of Fermi functions results in a step-like increase of
the current, when being overlapped with the transparency resonance
at the bias voltage equals a multiple of the phonon quantum
$n\Omega $, as shown in Fig. 1. Since the transparency doublets
are split faster than the Fermi windows are broadened out, after a
step-like increase the transmitted current slightly decreases with
the increasing voltage $V$, until the next vibrational quantum,
$n+1$, provides an additional channel for tunneling.

As the second example, we consider a model of the $C_{60}$ SET at
a large gate voltage $V_{gate}$ that stops the tunneling current.
The $V_{gate}$ eliminates the AOC and eliminates the charge
disturbances at the lead surfaces. Under these conditions, the
local electronic reservoirs remain at the equilibrium temperature
$k_BT\sim 0.5$ $meV$. If the Fermi distributions of electrons (see
Eq. \ref{Lan}) have sharp edges at $E=eV/2$, the step of the
current occurs when crossing the tunneling resonance $E=\Omega $
at $eV=2\Omega $, (see Fig. 2). The conductivity gap at double
frequency $2\Omega $ has been observed in Berkeley \cite{Park} by
measuring the differential conductance $\partial I/\partial V$ at
the frequency domain far from the bouncing-ball and the breath
modes of the $C_{60}$ transistor. The crossover regime to the
$2\Omega $ frequency has been found at large $V_{gate}$ that
suppresses the current through the $C_{60}$ transistor, as shown
in Fig. \ref{fig2} on the right hand side of the ${\partial
I}/{\partial V}$ plot \cite{Plot2d}. On the contrary, a small
$V_{gate}$ opens the transistor. Then, the current produces a
heating of the electronic reservoirs that smears the $2\Omega $
line.

The present scattering theory seems to work reasonably well, considering its
simplicity. The adiabatic picture explains the major features of the I-V
curves and especially their NDR behaviour, displayed in the Figs. \ref{fig1}
and \ref{fig2}. The bias voltage therein is below of the Coulomb energy, and
the charging of the quantum dot can be disregarded. But the broken Coulomb
blockade is the basic tenet of the electron transport, whose current growth
can be related to the electron affinity of quantum dot at $eV_{bias}\sim
U_c$. In the next section we demonstrate that the sequential tunneling can
affect the quantum dot paths by breaking symmetry of the adiabatic
potential.

\section{Charging and discharging in adiabatic theory of Coulomb blockade}

From the detailed balance we have learned that the charging is a
trade-off between the broadening $\Gamma $ and the equilibration
of the electron flows. The source-drain voltage can result in a
sudden change of this balance and the corresponding current.
Indeed, turning on the shuttle can maximize the transparency and
hence the conductivity. It can also affect the noise
characteristics. Qualitatively new effects such as low-frequency
coherent oscillations, electric echoes, and quantum entanglement
may arise due to coherence of the charge states.

In order to pick up a charge from the electronic terminal, several
mechanisms have to be involved in parallel. One- and many-body
effects (tunneling, screening, AOC etc.) are among them. Earlier,
considering the proximity surface effects, we have noticed that in
the equilibrium conditions the dot charge depends on the contact
potential. This is ensured by the electrochemical potentials $\mu
_\sigma $ $\sigma \in (l,r)$ and $\mu_{dot}$ in the leads and the
dot correspondingly, when these subsystems do not interact and can
be considered separately. The difference of the electrochemical
potentials creates a driving force for the electron flow, which
obeys the detailed balance conditions. The dot charges positively
or negatively, depending on whether $\mu _{dot}$ greater or less
than $\mu _\sigma $, and it remains neutral when $\mu _{dot}=\mu
_\sigma $.

Accordingly, the formulas for current and self-consistent charge
employ the electro-chemical potentials for the control of the
electron distributions in the leads and dot. The electro-chemical
potentials of the electrodes are defined as $\mu _\sigma =E_F\pm
eV/2$, where $E_F$ is the Fermi energy of electrons (including the
self-energy of pair interaction) and $V$ the external potential.
The electro-chemical potential $\mu _{dot}=U_{mean}(x,p)$ is
characterized, in turn, by the electron affinity given by the
charging potential $U_{mean}(x,p)$ so that the electronic energy
of the quantum dot is
\begin{equation}  \label{EnerInst}
E_{dot}=E(x,p)+U_{mean}(x,p),
\end{equation}
where $E(x,p)$ is the energy of the resonance electron level. For
the quantum dot between leads, the role of electron affinity and
ionization potential are played by the LUMO and HOMO state
energies, respectively. For the self-consistent electron
interaction, the mean-field potential $U_{mean}=U_c(Q-Q_0)$
denotes the charging (discharging) energy. The coefficient $U_c$
is the cost of Coulomb energy when the dot gains (loses) one
electron, $Q$ is the self-consistent charge and $Q_0$ is the
background charge, respectively. In the microscopic description,
calculation of the mean Coulomb energy relies on the Hartree-Fock
terms. The Coulomb energy integral depends on the wave-function
overlap accounting for the electron-electron interactions between
the electronic reservoirs and the dot.

This overlap measures the rate of tunnel transitions that entails
the detailed balance condition for the SETs. The transition
dynamics fall into the generic class of two-level-band systems
\cite{Akulin}. According to the common quantum-mechanical rules
\cite{Landay} we identify the following cases:

\begin{itemize}
\item Coherent tunneling via the virtual state of the dot: The resonance
level assists instantaneous transitions between electrodes, while the
resonance state remains empty.

\item Sequential tunneling via the real state of the dot: The dot charge is
not exhausted on the time scale of adiabatic dynamics
\footnote{e.g. by occasional hybridization with the surface
impurities.}. The charge transport could be driven by actual
shuttling.
\end{itemize}

Virtual and real tunneling transitions between the leads are
similar, in a sense, to Raman scattering and resonance
fluorescence for optical transitions. The critical dependence of
the shuttling on the bias voltage, that counteracts the relaxation
due to itinerant electrons and other dissipation mechanisms,
implies the importance of the latter for the onset of the
steady-state regime. This is similar to Vavilov's rule in optics,
stating that transitions through a real state are characterized by
their quenching ability. This analogy holds true for resonance
tunneling, in which the electron states and transitions are
treated quantum mechanically, while the adiabatic dynamics may
remain classical. The notion of the shuttle instability is related
to the dynamical symmetry of the bond potential (attractor) called
as the tunnel term, in which the dot resides. The critical
dependence of the tunnel term on bias voltage is demonstrated
below in the framework of a phenomenological approach. The quantum
description of the voltage-driven bifurcation is briefly sketched
in Sec. 7.

In the Breit-Wigner approximation, the charging - discharging
transitions are accounted for by the offset of the energy (or the
frequency shifts) dependence on the dot location relative to the
leads. According to Eq. \ref{Edot} it is
\begin{equation}  \label{e1}
E(x,p)=\varepsilon _0+\mathrm{E}_0(x,p)=\varepsilon
_0+K_{in}(p)+U_{pot}(x)+U_{ext}(x).
\end{equation}
The first part of the energy $\mathrm{E}_0(x,p)$ is the kinetic
energy $K_{in}(p)={p^2}/{2M}$. For the matter-field interaction,
this energy produces the well-known Doppler effect of
inhomogeneous broadening of optical transitions. The second part
is the interaction with electrodes, i.e. the Lennard-Jones
potential near the equilibrium point $x_0=0$ taken in the harmonic
approximation as $U_{pot}(x)={M\Omega ^2x^2}/{2}$. The third part
is the electrostatic interaction of the dot electron in the
external field between the electrodes
\[
U_{ext}(x)=-Fx.
\]
In order to formally compute the mean charge during the resonance
tunneling, we substitute the dot energy Eq. \ref{EnerInst},
\ref{e1} into the steady-state condition Eq. \ref{char}. This
yields
\begin{equation}  \label{char1}
Q(x,p)=\frac 1\pi \int\limits_0^\infty {dE\frac{{\Gamma
^rf_r(E)+\Gamma ^lf_l(E)}}{{(E-E_{dot}(x,p,Q))^2+\Gamma
^2}}}\mathop \cdot
\end{equation}
The Franck-Condon principle implies locality in phase space $x,p$.
It states that the coordinate and momentum of a massive dot stay
unchanged during the electron transitions, i.e. electron tunneling
in our case. For detailed balance, the tunnel coupling
$\Gamma_\sigma $ has to be invoked as the main prerequisite for
the Coulomb blockade. The additional relaxation due to molecular
collisions and electromagnetic radiation may have to be favored
for the steady state. This is similar to the dissipation
necessitated in the orthodox theory of the Coulomb blockade
\cite{Kulik}, \cite{Aver}.

The shuttling \cite{Gorel} also requires Coulomb blockade, which is a
combined effect of the single-electron charging and the quantized spectrum
of a bound system. The charging is produced by the bias voltage $V_{bias}$,
and hence it can be accompanied by a shuttle instability with threshold
behaviour. The physical meaning of the criticality is shown to be the Landau
bifurcation of the bonding potential. The vibrational energy, the
temperatures, and the charge dissipation prevent the shuttle mechanism
taking effect before the threshold $V_{bias}$ is reached.

\subsubsection{The self-consistent charge}

Mathematically speaking, we have arrived at a nonlinear
integro-functional equation, for which an exact analytical
solution is impossible, although its analysis within physically
meaningful approximations remains instructive. The integral over
the spectrum corresponds to the electron equilibrant flows ensued
by contact with two electron reservoirs. However, when the
inter-electrode distance is much longer than the Fermi length
$1/\gamma $ and the dot is trapped in the middle, $\gamma D\gg 1$,
the resonance window of the dot state density is extremely narrow:
$\Gamma (D/2)\approx E_Fe^{-\gamma D}\to 0$. Then, the integral
over the continuous spectrum of conducting electrons is collapsed
by the delta functions of the dot state density:
\begin{equation}  \label{QF}
Q(x,p)=\frac 1{\Gamma (x)}[\Gamma ^r(x)f_r(E_{dot})+\Gamma
^l(x)f_l(E_{dot})].
\end{equation}
In Sec. 3 we have already used this limit of narrow levels. The resonance
spectra of the electron tunneling without charging are explained by virtual
transitions. For real transitions through a charge state, the convoluted
integral \ref{char1} can be reduced to the nonlinear equation \ref{QF} for
the distribution $Q(x,p)$. The nonlinearity emerges from the condition of
detailed balance between the dot and the Fermi reservoirs $f_\sigma $, which
interact via irreversible coupling $\Gamma _\sigma $.

\begin{figure}[ht]
\sidebyside {\includegraphics[width=2.5in]{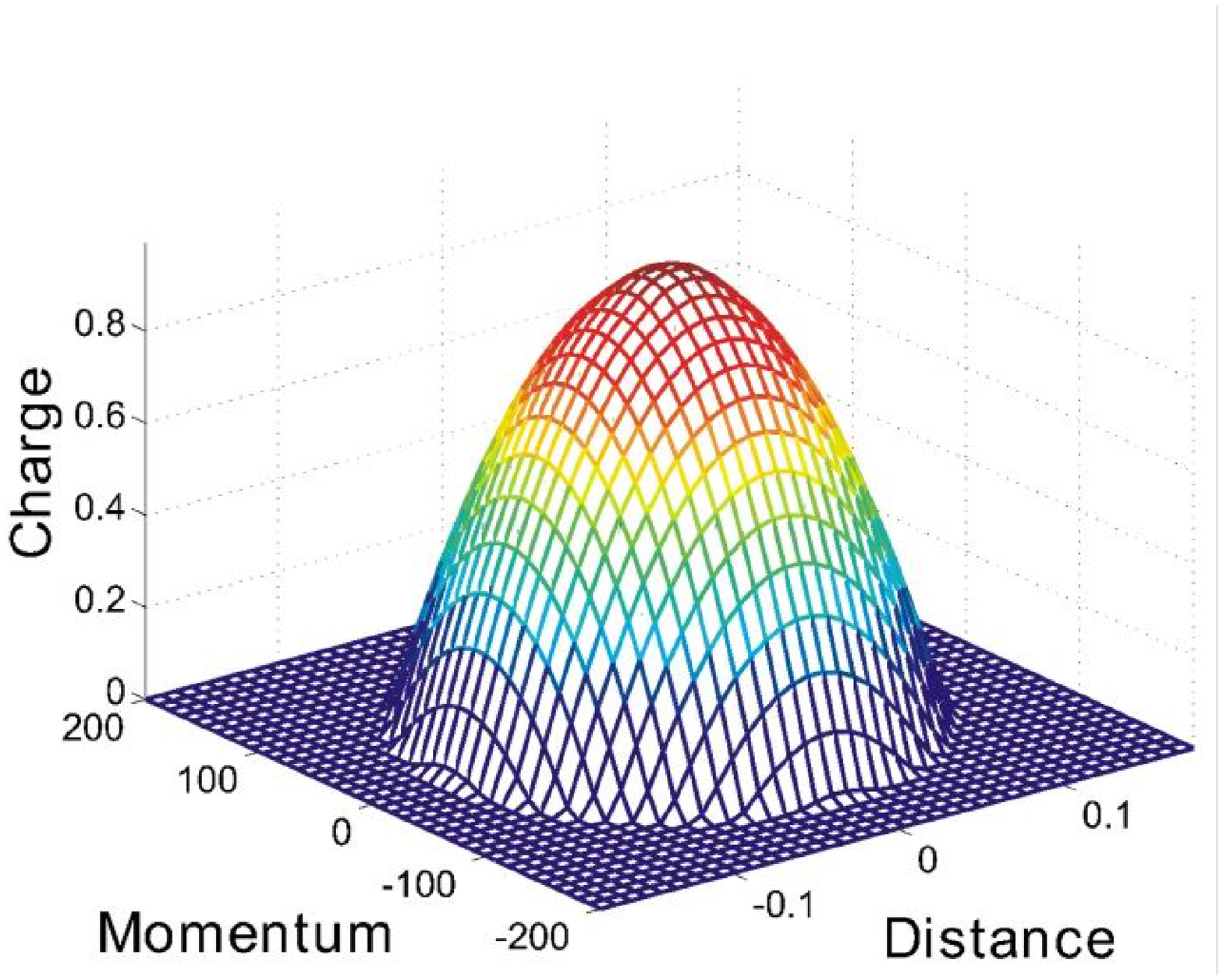}} {
\includegraphics[width=2.5in]{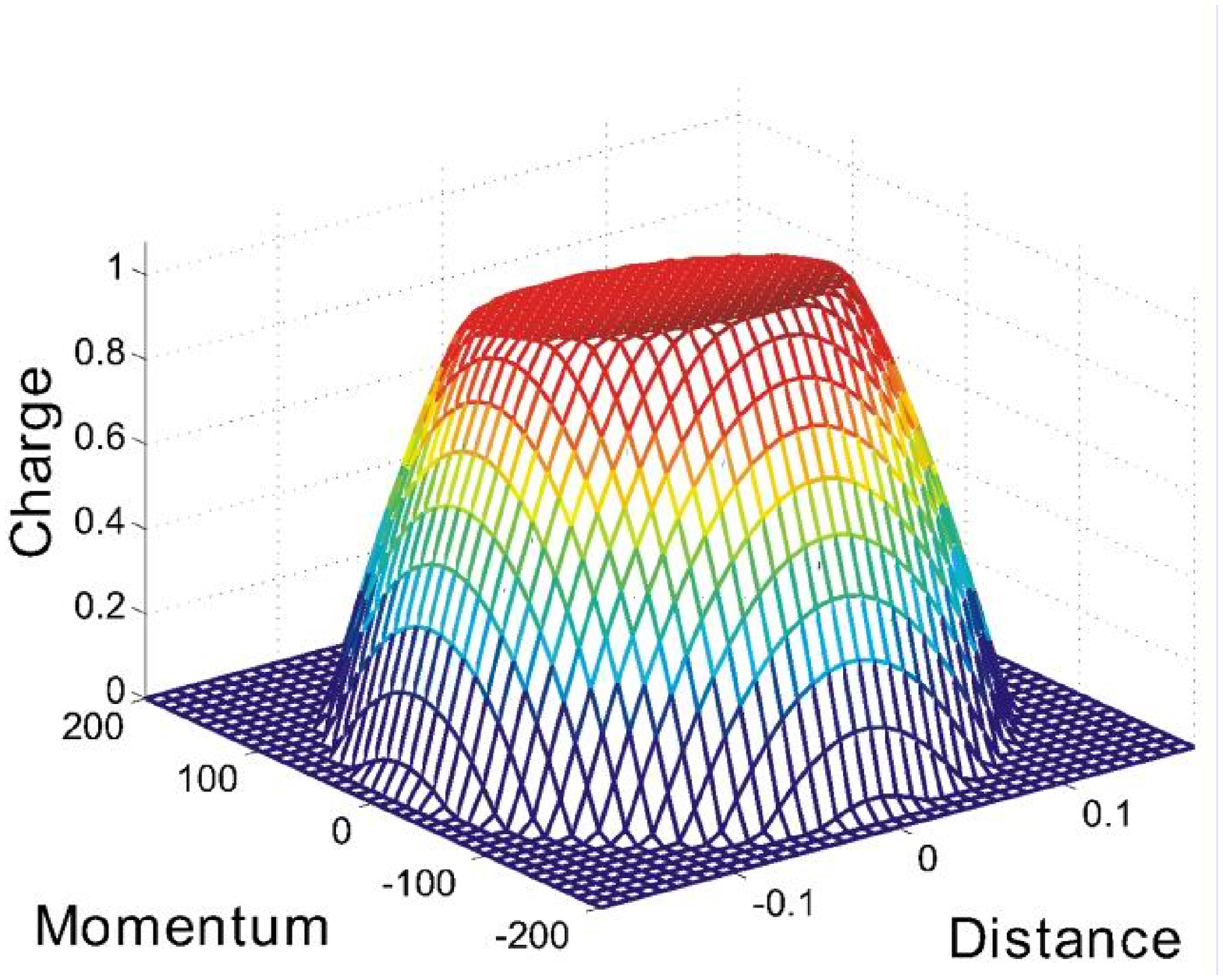}}
\sidebyside{\caption{The charge distribution in the $x$ and $p$
phase space of the vibrator with the zero point oscillation
amplitude $x_0=0.03$ and momentum $p_0=1/x_0$, acquires almost a
Gaussian shape for a bias voltage up to $V=100meV$.}\label{Char1}}
{\caption{On increasing voltage beyond $V=100meV$, the excess
charge grows from its background value to the saturated value
allowed by the Pauli principle. The saturation is characterized by
the voltage bias $V=150 meV$ which is accompanied by the triple
well potential.}\label{Char1a}}
\end{figure}

The mean charge $Q$ can be computed with the help of library routines, e.g.
by using either the Newton-Raphson algorithm or a globally convergent one.
The figures \ref{Char1}, \ref{Char1a} display distributions of the charge of
the molecular $C_{60}$ transistor, for which the thermal spread of the Fermi
functions is much larger than the tunnel broadening $kT_B\gg \Gamma $ at
bias voltages exceeding the Coulomb energy.
The charge yield (shown on the plots \ref{Char1}, \ref{Char1a}) is
related to the mean field that detunes the dot level from the
resonance. The detuning also includes the frequency shifts due to
the bonding interaction in the harmonic approximation near the
potential minimum. The use of the Lennard-Jones potential could be
made at longer distances, where the attraction to the surface is
created by the van der Waals or Casimir-Polder potentials. This
'spontaneous' interaction is relevant for the shuttle with a large
amplitude of vibrations close to electrodes' surfaces.

At moderate deviations from the equilibrium, we can chose the
adiabatic potential of the movable quantum dot as detailed above.
The first part is the potential energy $U_{pot}$ of the dispersion
interaction in the harmonic approximation. The second part
accounts for the electrostatic energy of charging. By definition,
it is the minimum energy required for adding charge $Q$ to the
quantum dot:
\begin{equation}
U_{adiab}(Q)=U_{pot}+U_{Q}+U_{ext},  \label{Uadiab}
\end{equation}
where the Coulomb and the external potentials are
$U_{Q}=U_{c}Q^{2}$ and $U_{ext}=-FxQ$, respectively. The
electro-chemical potential of the dot is the difference between
the adiabatic potentials of the adjacent charge states,
different by one electron 
\begin{eqnarray}
\mu _{dot} &=&U_{adiab}(Q+1)-U_{adiab}(Q)=\mathletter{a} \\
&=&2U_{c}Q-Fx+U_{c}.\mathletter{b}  \label{mudot}
\end{eqnarray}
The factor 2 in the electro-chemical potential of the quantum dot is to be
absorbed in the mean charge $Q$ , which accounts for the spin degeneracy in
the Eq. \ref{QF}. The energy $U_{c}$ shifts the resonance level. Without
loss of generality, we assume that the dot energy level $\varepsilon _{0}$
aligns with the Fermi energy of the  reservoirs. Then from Eq. \ref{mudot},
we obtain
\[
U_{mean}=U_{c}Q-Fx.
\]
for the self-consistent potential of the mean field. The adiabatic
potentials modify the conductivity (see Eq. \ref{Con}) in the
$x,p$ channel such that Landauer's formula relates the average
charge with the average current through the quantum dot. This
relationship is the central aim of any transport measurement
theory. Equation \ref{QF} achieves this aim in the mean field
approximation, in the spirit of the famous Weiss theory of
magnetic susceptibility.

Equation \ref{QF} has two mathematical properties important for the
adiabatic motion. Firstly, the dot charge in units $e$ has, for sure, to be
between zero and one. This circumstance helps one find an approximate
solution in the form of a converging power series over $Q$, provided
$u=\frac{U_{c}}{k_{B}T}\leq 1$. On the other hand, the decomposition over
exponent power $e^{uQ}$ can be employed for $u \ge 1$. Secondly, if the
thermal smearing is marginal, the dot population at the point $x,p$ of the
phase space switches quickly between the equilibrium states $0$ and $1$. The
solution is therefore expected to be expressed in terms of the step-like
Fermi functions or their derivatives.

One totally disregards the Coulomb blockade by neglecting the mean
field of charging in Eq. \ref{Uadiab}. In the zeroth
approximation, the $Q(x,p)$ is given by the sum of two Fermi
distributions in the reservoirs
\begin{equation}
Q^{(0)}(x,p)=\frac{1}{\Gamma }[\Gamma ^{r}(x)f_{r}(E_{0}(x,p))+\Gamma
(x)^{l}f_{l}(E_{0}(x,p))],  \label{Q0}
\end{equation}
each of which contributes proportionally to its coupling $\Gamma
^{\sigma }$ with the dot. Hence the charging becomes considerable
for $V\geq \Omega $, provided $V,\Omega \gg kT_{B}$. This means
that the magnitude of the bias voltage $eV$ has to be greater than
the zero-point energy of the dot vibrations (with amplitude
$x=x_{0}$ and $p=p_{0}$). Therefore
\begin{equation}
Q^{(0)}(x,p)\approx e^{\frac{eV/2-E_{0}(x,p)}{k_{B}T}}.  \label{Q00}
\end{equation}
\begin{figure}[th]
\caption{The tunnel terms at $\Omega =5meV$, $k_{B}T=0.5meV$,
$U_{c}=50meV$. The four curves correspond to
$V=0,25,50,75,100meV$. The thin lines refer to the exact numerical
solution, the dotted lines to \protect\ref{Q1} without the
logarithmic correction.} \label{Cba}\sidebyside
{\includegraphics[width=2.3in,height=2.22in]{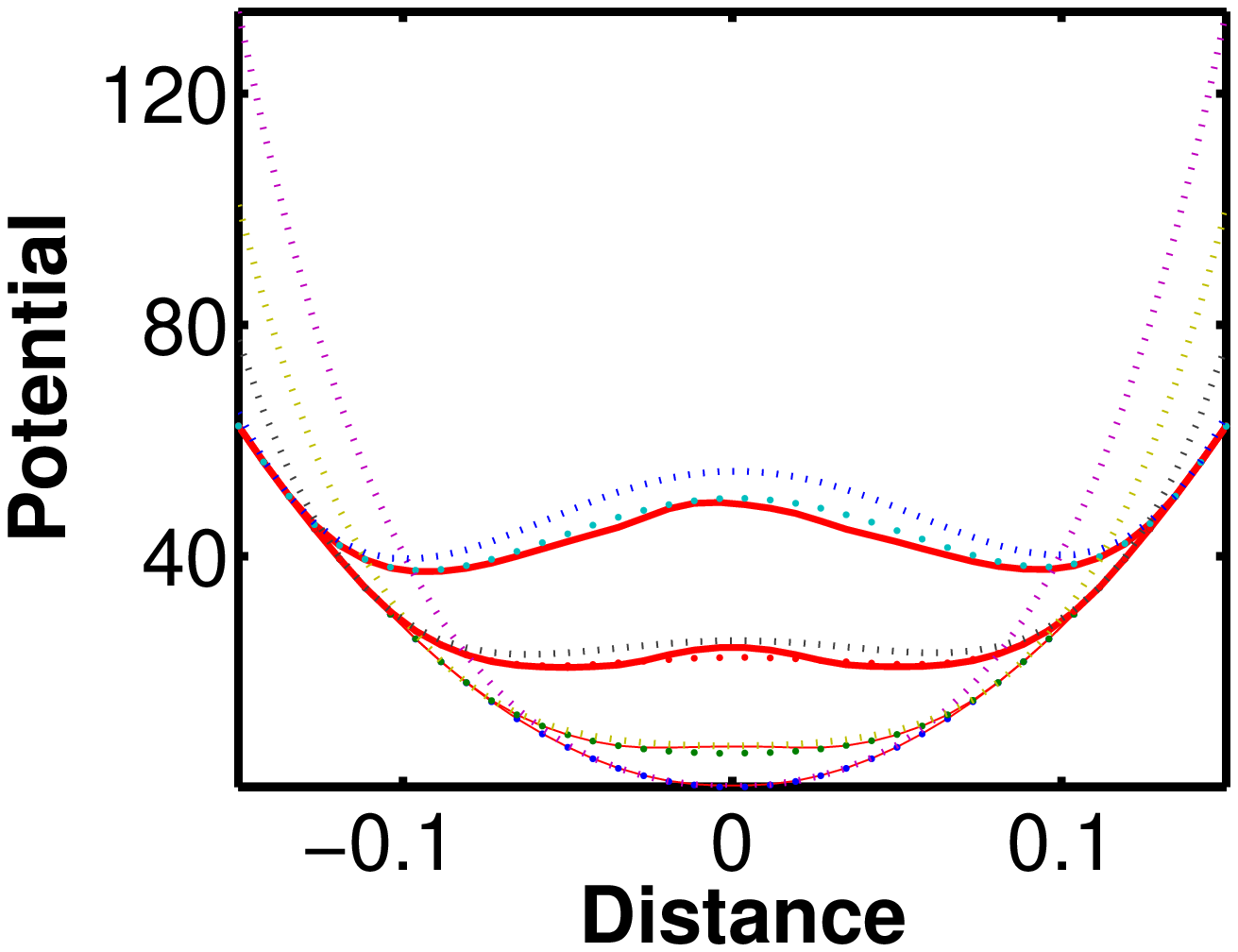}} {
\includegraphics[width=2.3in,height=2.2in]{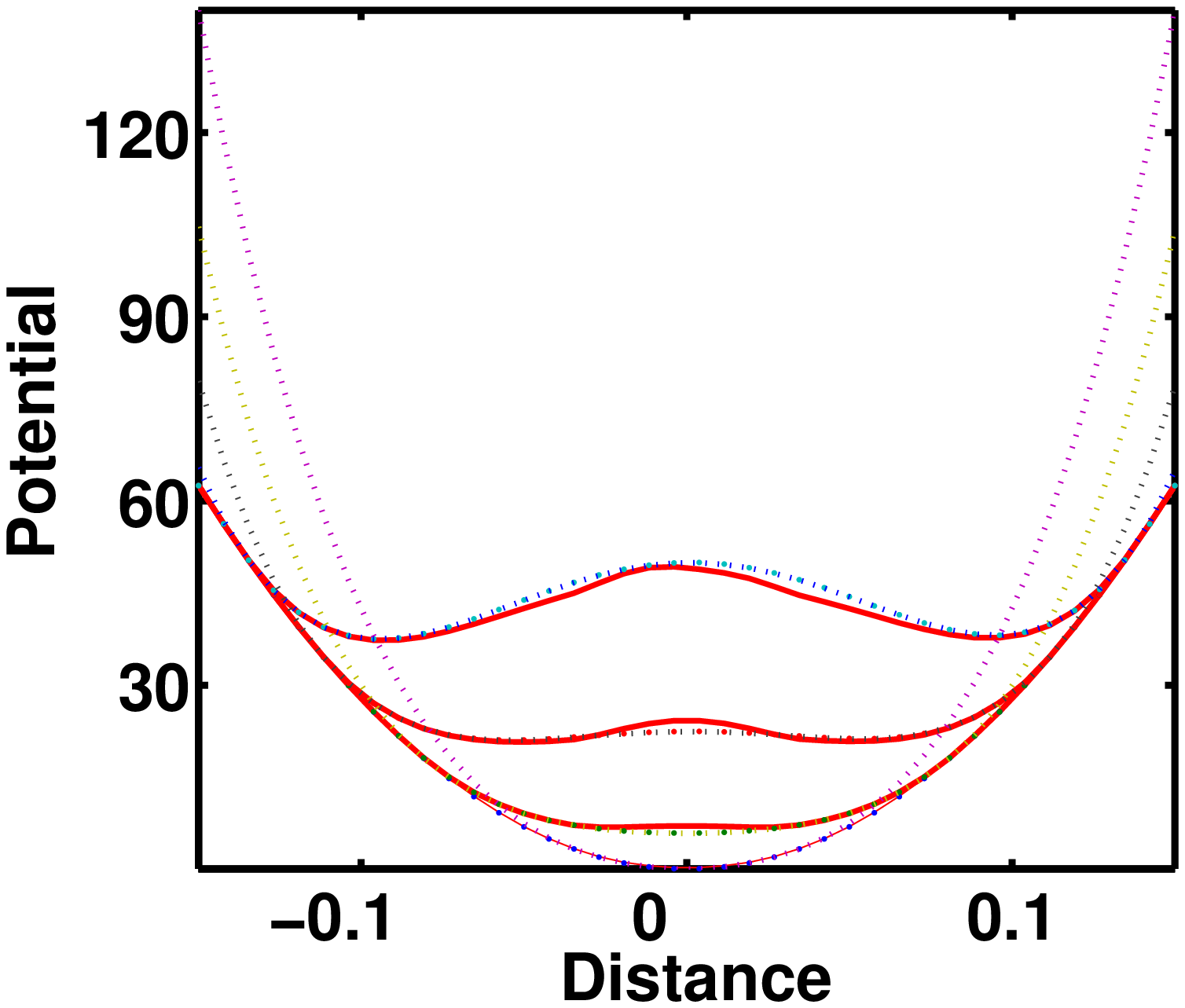}} \sidebyside {} {}
\caption{The same graphs but including the first logarithmic
corrections for Coulomb blockade. The points on the both plots
correspond to the numerical solution in the "Boltzmann" limit
where the saturation of the resonance level is
ignored.}\label{Cbb}
\end{figure}
This necessary condition corresponds to the onset of shuttling and has been
found in Ref.\cite{Fedor}. For $u \ll 1$, however, Eq. \ref{Q00} can be
utilized only at the beginning of the instability, because in this
approximation the charge is independent of $U_{c}$, which does not account
for the Coulomb blockade effect.

The molecular SETs are characterized by quite the opposite condition of
strong blockade, $u\gg 1$, where the Coulomb energy $U_{c}\sim
\frac{e^{2}}{4\pi r}$ is dominant. For instance, for $C_{60}$ molecular
radius $r$ $\sim 3.5$ $nm$, we obtain  that the Coulomb energy $U_{c}\sim
100$ $meV$ is larger than the vibrational energy $E_{0}\sim 5$ $meV $, and
the thermal spread $k_{B}T\sim 0.5$ $meV$. We have therefore to keep the
charge $Q$ in the exponent, instead of casting it into a power series. This
approximation is equivalent to the Boltzmann limit for the Fermi reservoirs,
where the saturation of population $f\sim 1$ is disregarded and the charge
distribution obeys the nonlinear equation
\begin{equation}
Q{(x,p)}^{{}}e^{\frac{{U_{c}Q(x,p)}}{{k_{B}T}}}=2\frac{{\cosh
(\frac{eV}{{2k_{B}T}}-2\gamma x)}}{{\cosh (2\gamma
x)}}e^{-\frac{{E_{0}(x,p)}}{{k_{B}T}} }.  \label{Bolt}
\end{equation}
An iterative solution can be obtained with the help of a
converging series of logarithmic corrections, the first two of
which read
\begin{equation}
Q(x,p)\approx
{\frac{eV-2E_{0}(x,p)}{2U_{c}}}-\frac{k_{B}T}{U_{c}}\ln {\frac{
eV-2E_{0}(x,p)}{2U_{c}}},  \label{Q1}
\end{equation}
being derived from Eq. \ref{Bolt} under the typical conditions of molecular
SET $eV\gg k_{B}T$, $\gamma x\ll 1$. Comparison of the approximative
solutions with the exact numerical calculations of Eqs. \ref{Bolt} and
\ref{QF}, shown in Figs. \ref{Cba} and \ref{Cbb}, demonstrates the good
accuracy of the Boltzmann limit. This concerns the logarithmic corrections
Eq. \ref{Q1} as well, until the bias voltage reaches a value of the order of
the Coulomb energy, and $E_{0}\sim eV/2$.

\subsubsection{The adiabatic Hamiltonian}

\begin{figure}[th]
\caption{The charge distribution (at the zero $p$ slice) at the bias
voltages at $V=0,25,50,75,100$ meV}. \label{BSCa}\sidebyside
{\includegraphics[width=2.5in]{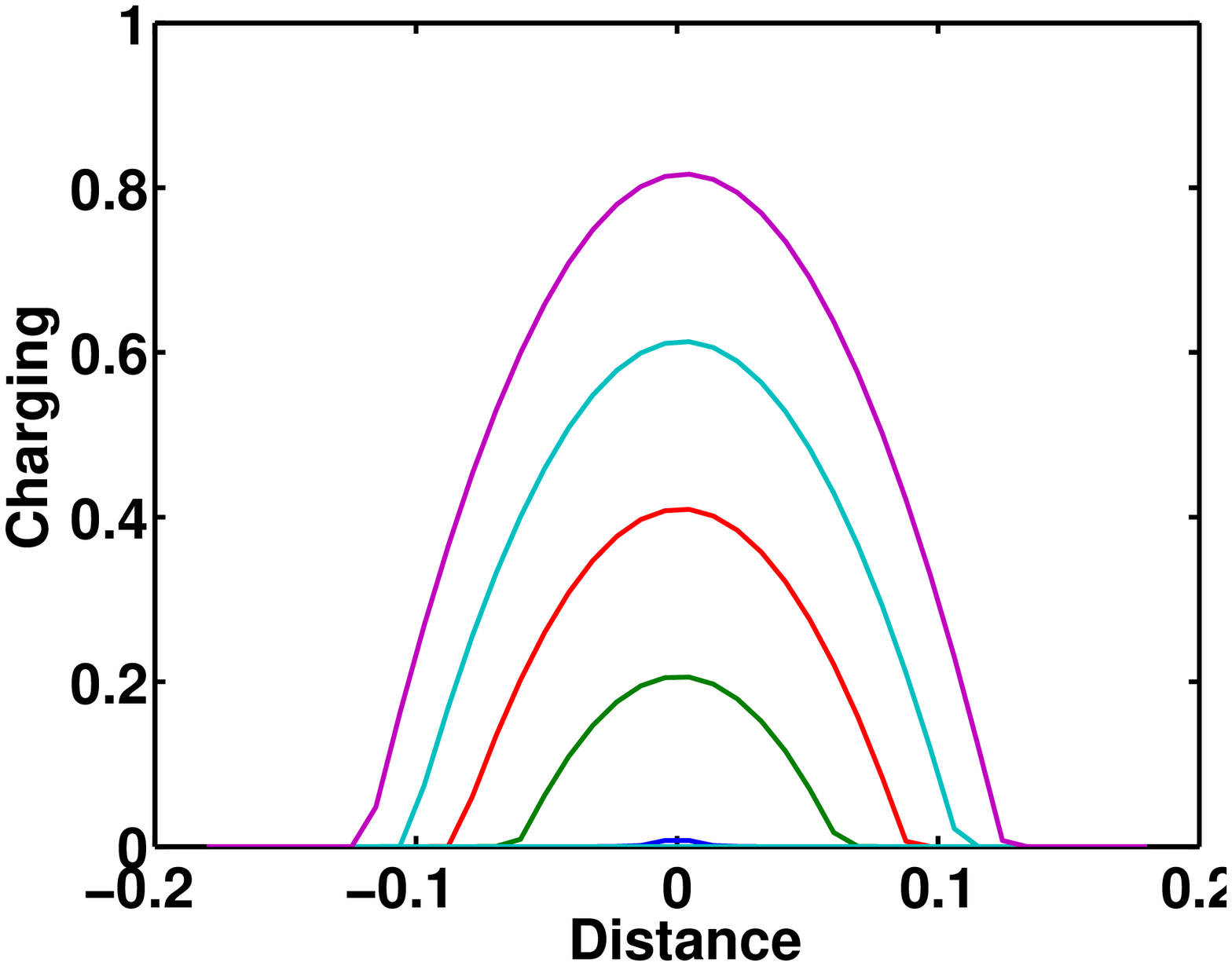}} {
\includegraphics[width=2.5in]{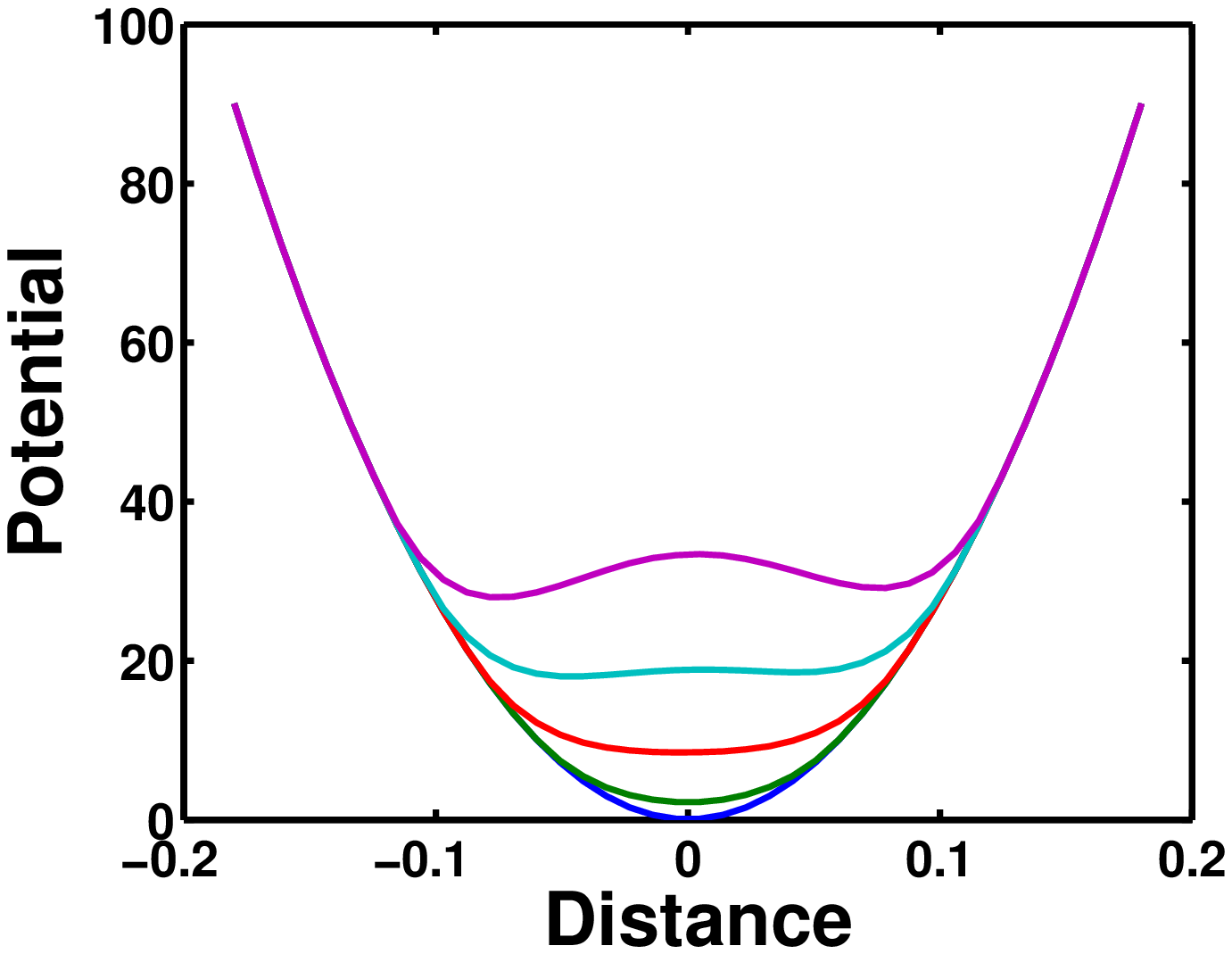}} \sidebyside {}{}
\caption{The adiabatic potential $U_{c}(Q)$ at the same voltages $\Omega
=5meV$, $k_{B}T=0.4meV$, and $U_{c}=50meV$.}\label{BSCb}
\end{figure}
Dynamics of mechanical motion of the dot in $x,p$ phase space is
governed by the Hamiltonian
\begin{equation}
H(x,p)=K_{in}(p)+U_{adiab}(Q)=\frac{1}{2M}p^{2}+\frac{M\Omega ^{2}}{2}%
x^{2}+U_{c}Q^{2}-FxQ.  \label{AH}
\end{equation}%
By substituting the eq. \ref{Q1}  into eq. \ref{AH} and omitting the small
logarithmic correction one obtains the adiabatic potential
\begin{equation}
U_{adiab}(x)=\frac{M\Omega
^{2}}{2}(1-\frac{eV}{U_{c}})x^{2}-\frac{Fx}{2U_{c}}(eV-M\Omega
^{2}x^{2})+\frac{M^{2}\Omega ^{4}}{4U_{c}}x^{4}  \label{AP0}
\end{equation}
for $p=0$, which has either a single- or double-well shape,
depending on the bias voltage. The threshold voltage, at which the
frequency of vibrating dot becomes zero, coincides with the
Coulomb energy $U_{c}$.  At this threshold, the original solution
of the classical equations of motion near the potential energy
minimum at the point $x=0$ becomes instable. The bond symmetry is
therefore broken by the bias voltage due to the dot charging and
the Coulomb blockade. The single-well potential is hence
transformed into a double-well potential with new stable points
locating far from the original point of equilibrium. This broken
symmetry is shown in Figs. \ref{BSCa} and \ref{BSCb}, obtained
from the exact nonlinear Eq. \ref{QF}. The figures \ref{Cba},
\ref{Cbb} also illustrate this effect in the Boltzmann limit of
Eq. \ref{Bolt}, where the analytical solution of Eq. \ref{Q1} is
compared with the logarithmic approximation.

\subsubsection{Tunnel terms symmetry and phase space structure of the dot
motions}

\begin{figure}[th]
\sidebyside {\includegraphics[width=2.5in]{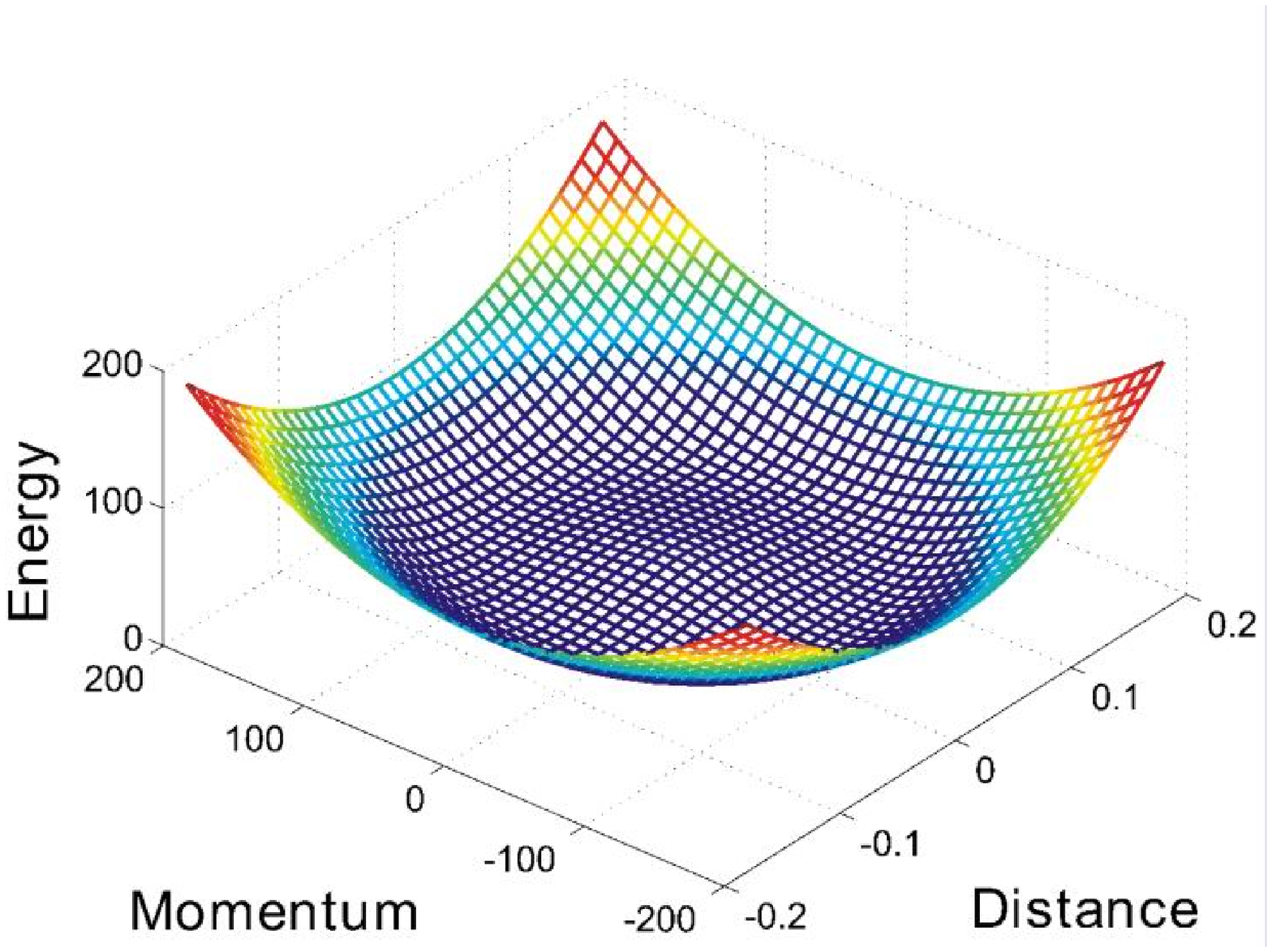}} {
\includegraphics[width=2.5in]{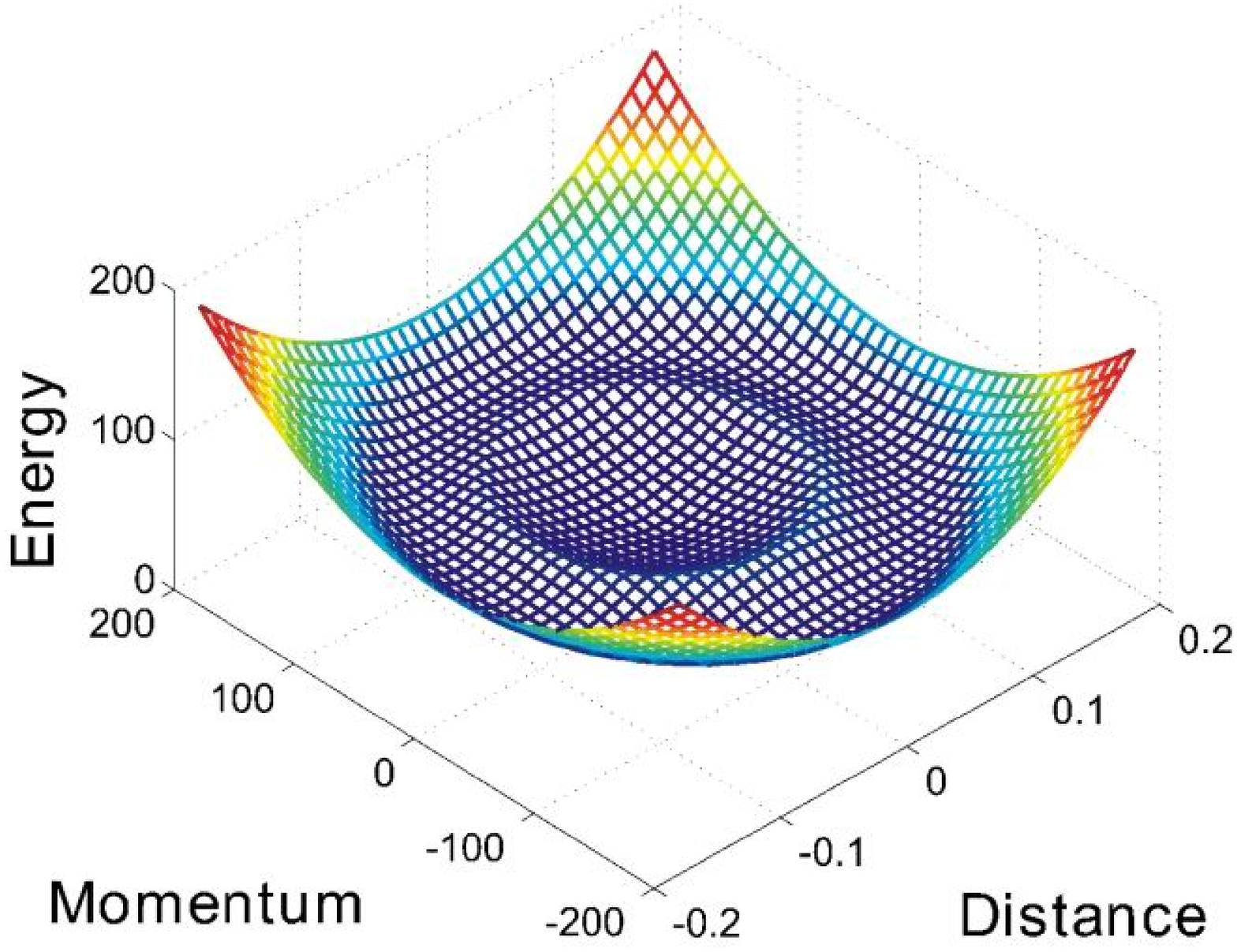}}
\sidebyside{\caption{The energy functional $K_{in}(p)+U_{adiab}(Q)$
bifurcates into the double well potential at $eV=100$ meV. The barrier
between wells is caused by the finite cost of the Coulomb energy $U_c=50$
meV.}\label{Char2}} {\caption{The voltage bias growth causes the double well
to resolve into the triple well at the second threshold when the resonance
state population is saturated. The saturation depends on the temperature and
the Coulomb energy.}\label{Char2a}}
\end{figure}

Now we are in a position to discuss the classical trajectories of the
quantum dot driven by the bias voltage. To this end, the charge
distributions $Q(x,p)$, plotted in Figs. \ref{Char1} and \ref{Char1a} are
substituted into Eq. \ref{AH}, and the resulting adiabatic Hamiltonians in
phase space are shown in Figs. \ref{Char2}, \ref{Char2a}.  In the
thermodynamic limit, the dot tends to occupy the minimum-energy valley due
to irreversible relaxation. The valley topology strictly depends on the bias
voltage. The dot trajectories are running towards the center of the circle,
$x=0$ and $p=0$, until the first threshold is reached.
The source-drain voltage specified below breaks the symmetry of the single
well potential (for the case of the ground state frequency $\Omega =5meV$ it
is shown in Fig. \ref{Char2}). The center of attraction at the point $x=0$,
$p=0$ becomes unstable: The fixed points settle outside the circle, where
the dot trajectories are winding. The initially compact dot distribution in
the $x,p$ phase space plane spreads into a crescent-shaped 'moon', and, then
into a ring distribution of a radius found from Eq. \ref{AP0} by setting
\[
\frac{{\partial U_{adiab}(Q)}}{{\partial x}}=0.
\]
If the bias voltage is larger than the Coulomb energy $U_{c}$, this
condition yields new stationary points of equilibrium
\begin{equation}
x^{2}=\frac{V-U_{c}}{2M\Omega ^{2}} \mathop ,   \label{AP1}
\end{equation}
corresponding to the minimum potential energy $U_{adiab}(Q)$. Furthermore,
with the increase of the bias voltage the bond symmetry is broken again. The
persistent Coulomb blockade produces the triple-well potential by combining
 the single and double wells into unified potentials as shown in Figs.
 \ref{Char2} and \ref{Char2a}. The multiple-well structures can not be
inessential for the quantum dynamics. The coherent tunneling of
electrons entails coherence between the wells, as will be
discussed in Sec. 9. However, before plunging into the quantum
theory we wish to turn our attention to the classical aspects of
the system noise.

\section{The shot noise of the shuttling instability}

At the voltage threshold $V\sim \Omega $ the dot begins to
oscillate with a larger amplitude between electrodes thus
augmenting the current and its fluctuations. For the molecular
SET, an explanation of the current-voltage steps has been
attempted in terms of the Frank - Condon transitions \cite{Bose}.
The shuttling mechanism in the coherent regime \cite{Fedor} has
been advocated in the works of Shekhter's group in Chalmer's
\cite{Gorel}. However, it seems that  the kinetic models of
incoherent tunneling offer the best fitting \cite{Braig} of the
current-voltage curves for the $C_{60}$ SET \cite{Park}.

To resolve the dilemma, one needs an additional analysis of the
conduction mechanisms. For instance, the amount of electrons
transported per period has been estimated on the basis of the
known source-drain current $I\sim 100pA$ \cite{Park}. However,
when taking into account the frequency $\Omega \sim 1.2THz$ of the
"bouncing-ball mode", it turned out that the $C_{60}$ transistor
transmits only a small fraction $q\sim 10^{-3}e$ of the elementary
charge $e$ per vibrational period $\tau =2\pi /\Omega \sim
10^{-12}sec$. This excess charge is too small to compete with and
to affect the shuttling dynamics of the $C_{60}$ SET.

In principle, an objection based on the broken bond symmetry model
can be raised, because the frequency of dot vibrations is softened
to zero at the onset of the shuttling instability. However, as
shown above, e.g.  \ref{AP1}, this could occur at a bias voltage
of the order of the Coulomb energy $V\sim U_{c}\gg \Omega $, which
is considerably larger than the observed conductivity gap $\Omega
$ \cite{Park}. The current by itself cannot instruct us
unambiguously about its transport mechanisms. This suggests one
should try more advanced tools dealing with fluctuations rather
then just with the mean current. For instance, the noise spectrum
or a more general full-counting statistics could reveal more
subtleties. In particular, this also concerns the Fano factor,
which measures the ratio of the actual noise of fluctuation
spectrum and the Poisson shot noise produced due to the tunneling
of isolated independent electrons
\[
F_{ano}=\frac{S_{noise}}{S_{Schottky}}.
\]
This Poissonian limit, also termed as Schottky noise, has been
obtained for a very low transparency $\Upsilon \ll 1$ and is
proportional to the conductivity Eq. \ref{Cond}. We consider the
resonance case, for which the dot oscillates on an adiabatic
trajectory $x=A\cos (\Omega t)$ with the transparency
\begin{equation}
\Upsilon (\varphi )\sim \frac{\Gamma ^{r}(\varphi )\Gamma
^{l}(\varphi )}{\Gamma (\varphi )}=\frac{\Gamma _{0}}{\cosh
(2A\gamma \cos (\varphi ))},
\end{equation}
where $\phi =\Omega t$. The Schottky noise is proportional to the mean
current (see Eq. \ref{noise}) averaged over the dot trajectory:
\begin{equation}
\langle I\rangle =\frac{{e^{2}V}}{h}\int\limits_{0}^{2\pi
}{\frac{{d\varphi }}{{2\pi }}\Upsilon (\varphi )}.
\end{equation}
The main contribution to this integral comes from the neighborhood of $\phi
=\pi /2$ where the transparency is maximum, and the noise is minimum
according to
\begin{equation}
S=\frac{{2e^{3}V}}{h}\int\limits_{0}^{2\pi }{\frac{{d\varphi
}}{{2\pi }} \Upsilon (\varphi )(1-\Upsilon (\varphi ))}.
\end{equation}
\begin{figure}[tbp]
{\includegraphics[width=3in, height=3in]{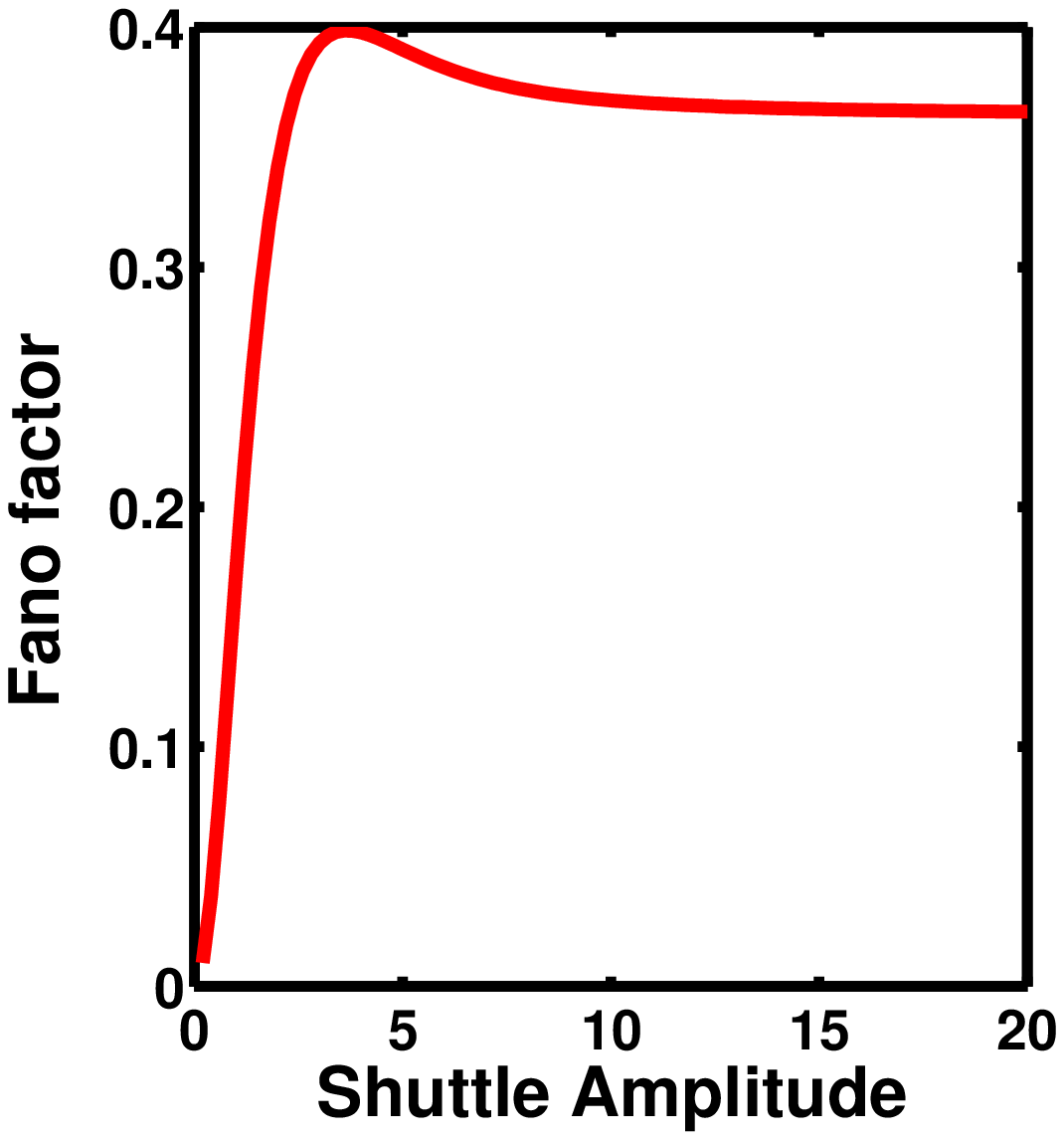}}
\narrowcaption{The Fano factor as function of  amplitude  $A$ of
the dot oscillation. The static dot ($A=0$) has the resonance
transparency $\Upsilon=1$ and is characterized by zero noise. The
$F_{ano}$-factor grows rapidly until $A$ reaches the threshold
value of $A_{thr}\sim \gamma^{-1}$ the Fermi length. After that
the shuttle mechanism dominates, enhancing  the transparency and
correspondingly decreasing the noise slightly to keep it below of
the Poissonian limit.} \label{Fano}
\end{figure}
The Fano factor $F_{ano}$, (see figure \ref{Fano}) as a function
of the shuttle amplitude is not a monotonic curve. At rest, $A=0$,
the noise is minimum for resonance tunneling since the dot energy
level is not broadened and the transparency is ideal $\Upsilon
=1$. With increasing $A$ the $F$-factor rapidly grows towards the
threshold value $A\sim 1/\gamma $, and thereafter one is in the
shuttling regime. The electron transport favors a decrease in the
$F$-factor and augments the noise amplitude $A$ by a value larger
than the Fermi length $\lambda \sim 1/\gamma \sim 1\mathring{A}$.
Figure \ref{Fano} shows the Fano factor as a function of shuttling
amplitude and illustrates that the better the transparency the
lower the noise, as it should be for a quiet electronic sea.

This analysis shows that shuttling is hardly feasible in the
reported experiments \cite{Park}. The first reason is the AOC that
is caused by the tunneling at a small gate voltage. Growth of the
local environment temperature destroys the Coulomb blockade, which
could lead to charge accumulation and shuttled transport. The
second reason is a large gating of the $C_{60}$ transistor that
removes the AOC and thermalizes the SET at low temperature
$T\sim4K$. Nevertheless, the amplitude of the thermal vibration is
still too small to activate the shuttle mechanism at distances
comparable with the Fermi length $\lambda \sim 1\mathring{A}$. The
threshold of shuttling $A_{thr}\sim \lambda $ can not be reached
even at $eV\sim \Omega =5meV$ for the $C_{60}$ transistor.

If zero-point vibration amplitudes of the dot are comparable with
the Fermi length of the electrons, the shuttling takes place at
small bias voltage. This is the case for cold dots. The
constructive interference of electron waves in the tunnel gap
center effectively charges the dot.  In the quantum limit, this
charging requires a justification of the tunnel-term concept based
on the Schr\"{o}dinger equation. In next section we address a more
rigorous quantum mechanical picture based on the "ab-initio" SET
model.

\section{The Born-Oppenheimer approximation for the tunnel curves}

In previous section,  by considering the electrostatic energy of
the quantum dot charging we have determined the tunnel curves
using the phenomenological approach. A strict definition of the
tunnel curves as total electronic energy at a fixed dot location
between leads is implied by the Born-Oppenheimer adiabatic
strategy. For the quantum-mechanical computation of the tunnel
curves, the information about   (1) the spatial profile of
electrostatic potential and (2) the electron and ion distributions
of the SET is required as an input.

The electrostatic interaction of the quantum dot with the leads is
not weak due to charge images. Metallic leads screen their Coulomb
interaction by creating charge holes in the electron reservoirs
and by breaking the translation symmetry of the electron-electron
and the electron-hole interactions. The quantum dot is subject to
the dipole and multipole forces, which derive from the tunnel
terms. Besides that, the screening diminishes the threshold of
cold emission. The tunneling rate increases by three-orders of
magnitude \cite{Flugge} as a result. We therefore have to consider
the quantum dot and the leads as a strongly interacting combined
system when computing the resonance tunneling as a function of the
dot position in the tunnel gap.

The Green's function of the electrostatic field inside the double junction
obeys the 3D-Poisson equation:
\begin{equation}
(\partial _{x}^{2}+\partial _{y}^{2}+\partial
_{z}^{2})G(x,y,z,x_{1},y_{1},z_{1})=\delta (x-x_{1})\delta (y-y_{1})\delta
(z-z_{1}).  \label{PE}
\end{equation}
The metallic electrodes are implied to be equipotential surfaces at which
the boundary conditions read
\begin{equation}
G(\pm D/2,y,z,\pm D/2,y_{1},z_{1})=0,
\end{equation}%
for all points $y,z,y_{1},z_{1}$. The solution can be represented
as a sum of two components. The first one satisfies the
homogeneous Poisson equation with the boundary condition provided
by the voltage applied. This part gives us an external potential
resulting in a Stark shift. The second part is responsible for the
charging and for the Coulomb electron-electron interaction. The
potential field of the charge obeys the Poisson equation, which
has a delta-source on the right hand side of Eq.\ref{PE} at zero
boundary conditions. The solution of the inhomogeneous equation is
given by a two-fold Fourier integral. The transversal isotropy of
the tunneling electron flow allows one to reduce it to a one-fold
integral
\begin{equation}
\varphi (\chi ,\chi _{1})=4e^{2}\int\limits_{0}^{\infty }{d\lambda
}\left[ {\frac{{J_{1}(\lambda a)}}{{\lambda a}}}\right]
^{2}\frac{{sh[\lambda (\frac{D}{2}+\chi _{<})]\mathop{}sh[\lambda
(\frac{D}{2}-\chi _{>})]}}{{sh(\lambda D)}},
\end{equation}
which describes the field between the tips with a charge located
at the point $\chi _{1}$, where $\chi _{>}=\max (\chi ,\chi
_{1}),_{{}}\chi _{<}=\min (\chi ,\chi _{1})$. Here $J_{1}(a)$ is
the Bessel function, and the tunneling radius ${a}$ equals the
molecular van der Waals radius of the dot. The electron tunneling
from  the electrode tips occurs uniformly over a sphere of this
radius.
\begin{figure}[th]
\caption{The typical tunnel curve of the dipole dot as function of its
coordinate. The source-drain voltage bifurcates at a threshold voltage: a
single well is replaced by a double well followed by a wide well.}
\label{tuncurve}\sidebyside {\includegraphics[width=2.4in]{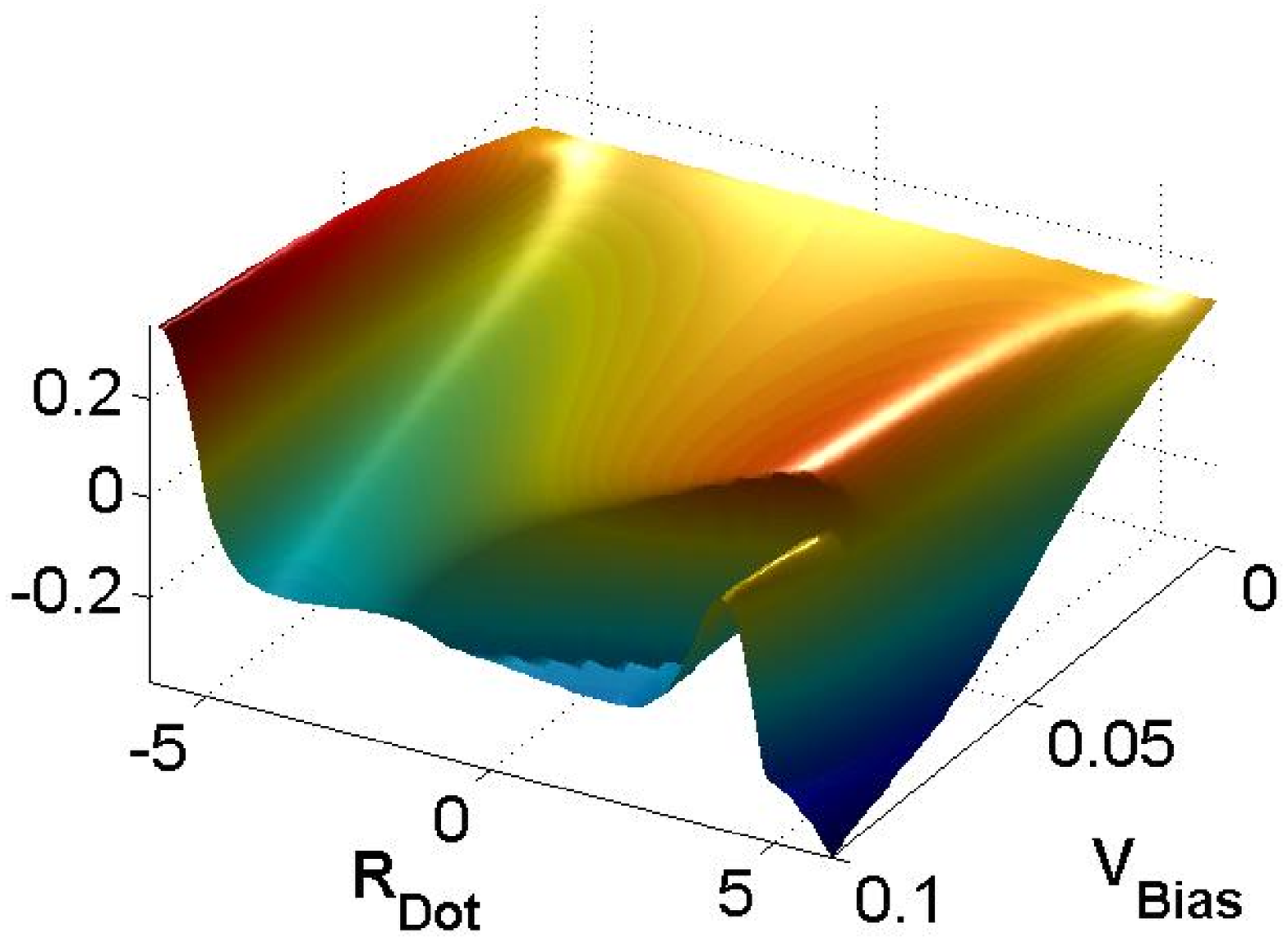}} {
\includegraphics[width=2.4in]{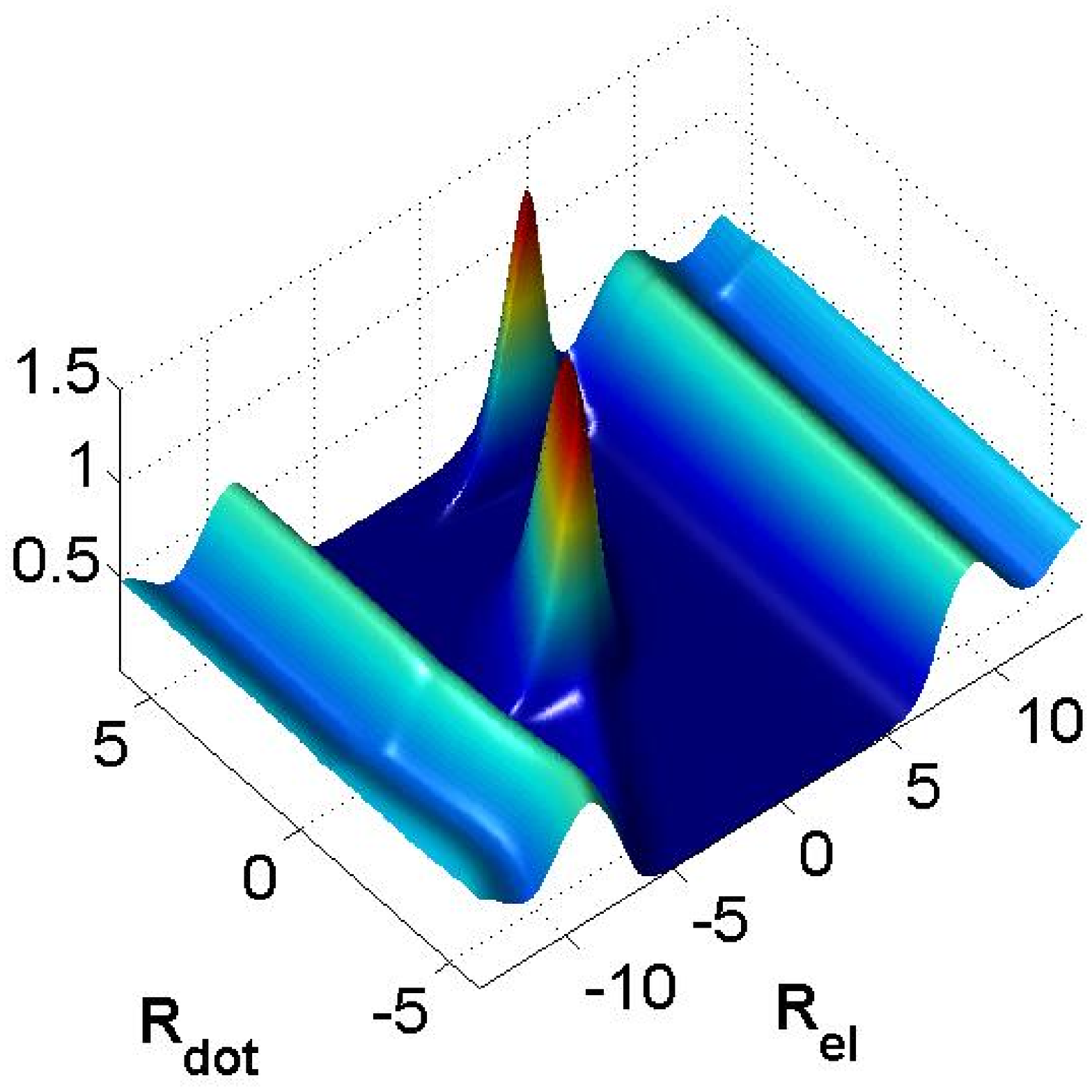}} \sidebyside {} {}
\caption{The electron density distribution inside SET displays Friedel
oscillations at the Fermi edges and the peak of resonance tunneling in
classically forbidden region in the inter-electrode gap. }\label{edm}
\end{figure}
The charge density of the dipole of the length ${a}$ is
\begin{equation}
n_{d}(\chi ,x)=F\left( {\delta (\chi -x+a)-\delta (\chi -x-a)}\right) ;
\end{equation}%
where the $F=\alpha V/D$ and $\alpha $ is the dot polarizability. The
electronic density $n_{e}(\chi ,x)$ of scattering waves can be found from
the wave function $\psi _{E,\sigma }(x)$ constructed from the transmitted
and the reflected plane waves
\begin{eqnarray}
\psi _{k,l}(x<-\frac{D}{2})=
&&e^{ik(x+\frac{D}{2})}+R_{k,l}e^{-ik(x+\frac{D}{2})}, \\
\psi _{k,r}(x>+\frac{D}{2})= &&e^{-ik_{r}(x-\frac{D}{2}
)}+R_{k,r}e^{+ik_{r}(x-\frac{D}{2})}, \\
\psi _{k,l}(x>+\frac{D}{2})= &&T_{k,l}e^{+ik_{l}(x-\frac{D}{2})}, \\
\psi _{k,r}(x<-\frac{D}{2})= &&T_{k,r}e^{-ik(x+\frac{D}{2})},
\end{eqnarray}
where the partial channels $k_{l}=k_{r}=\sqrt{2E+eV}$ and $k=\sqrt{2E}$ are
averaged over the thermal reservoirs of the leads. This yields
\begin{equation}
n_{e}(\chi ,x)=\sum\limits_{\sigma
=r,l}{\int\limits_{0}^{\infty}{dE} f_{\sigma }(E)\left\vert {\psi
_{E,\sigma }(\chi ,x)}\right\vert ^{2}}.
\end{equation}
The net charge distribution
$\rho(\chi,x)=n_{e}(\chi,x)-n_{d}(\chi,x)$ can be used for
computing the energy functional of the electron-electron and
electron-hole interactions in the tunnel gap. The potential energy
of the induced charges as a function of the dot location is
proportional to the integral
\begin{equation}
P_{H}(x)=\frac{1}{2}\int\limits_{-D/2}^{D/2}{d\chi
}\int\limits_{-D/2}^{D/2}{ d\chi _{1}}\mathop{}\rho (\chi
,x)_{{}}\varphi (\chi ,\chi _{1})_{{}}\rho (\chi _{1},x),
\label{tunterm}
\end{equation}
which measures the electron and hole overlap. It is related to the
rate of the tunnel transitions by the detailed balance condition
between the Fermi reservoirs and the dot. The kinetic energy also
depends on the dot location and can be calculated as
\begin{equation}
K_{el}(x)=\sum\limits_{\sigma =r,l}{\int\limits_{0}^{\infty
}{\frac{dE}{2}} f_{\sigma }(E)\int\limits_{-D/2}^{D/2}d\chi
\left\vert \frac{\partial \psi _{E,\sigma }(\chi ,x)}{\partial
\chi }\right\vert ^{2}}.
\end{equation}
Both the electrostatic and the kinetic parts combine into the
total electron energy. As a function of the coordinate $x$, this
energy plays the role of a mechanical potential
\begin{equation}
U_{T}(x)=\frac{K_{el}(x)+P_{H}(x)}{\int\limits_{-D/2}^{D/2}d\chi
\mathop{} n_{e}(\chi ,x)}
\end{equation}
of the quantum dot in the classically forbidden region of the
electron motion. This adiabatic potential is called a tunnel
curve, by analogy to molecular curves in the Born-Oppenheimer
theory. The tunnel curve depends on the location of the quantum
dot in the tunnel gap and on the Fermi distribution of the
electronic reservoirs. The electron density in the quantum dot is
obtained by solving  the Hartree equation for the electron wave
function
\begin{equation}
-\frac{1}{2}\frac{\partial ^{2}\psi _{E,\sigma }(\chi ,x)}{{\partial \chi
^{2}}}+[U_{ext}(\chi )+U_{H}(\chi ,x)]\psi _{E,\sigma }(\chi ,x)=E\psi
_{E,\sigma }(\chi ,x)
\end{equation}
with the boundary conditions at the electronic reservoirs, where
the potential $U_{H}(\chi ,x)$ reads
\begin{equation}
U_{H}(\chi ,x)=\int\limits_{-D/2}^{D/2}{d\chi _{1}}\varphi (\chi ,\chi
_{1})\rho (\chi _{1},x).
\end{equation}
One should also invoke Fermi statistics. A typical tunnel curve is
shown in Fig. \ref{tuncurve} for SET model with
$D=14\mathop{}[au]$, $a=1 \mathop{}[au]$, the work function of
electrodes $W=0.4\mathop{}[au]$, the Fermi energy
$E_{F}=0.2\mathop{}[au]$, and the polarizability $\alpha =200
\mathop{}[au]$ (of Na atom). The potential drops near the
interface of the source-drain electrodes, as it should for the
ballistic regime. The tunnel curve has a single shallow well at a
small bias voltage. When the latter increases, the well becomes
deeper, and the dot is attracted to the inter-electrode gap center
due to resonant tunneling. When the bias voltage grows further,
the second well appears near the drain electrode. Then, the wells
are broadened and combined into a single wide potential thus
giving rise to the shuttle mechanism of conductivity, as displayed
in Fig. \ref{tuncurve}. This figure justifies, in principle, the
broken-symmetry mechanism discussed in Sec. 5 in the framework of
a phenomenological model. The electron density as a function of
the electron and the dot coordinates is plotted in Fig. \ref{edm}.

\subsubsection{Density of states and current simulation}

The edge electron density is subject to Friedel oscillations due
to interference between the incident and the reflected electron
waves.
\begin{figure}[ht]
\caption{The current through the dipole as function of its
position inside the tunnel gap and bias voltage. The maximum
tunneling current is concentrated at the gap center, where
constructive interference of electron flows takes place.}
\label{fig10}\sidebyside {\includegraphics[width=2.4in,height=2.4in]{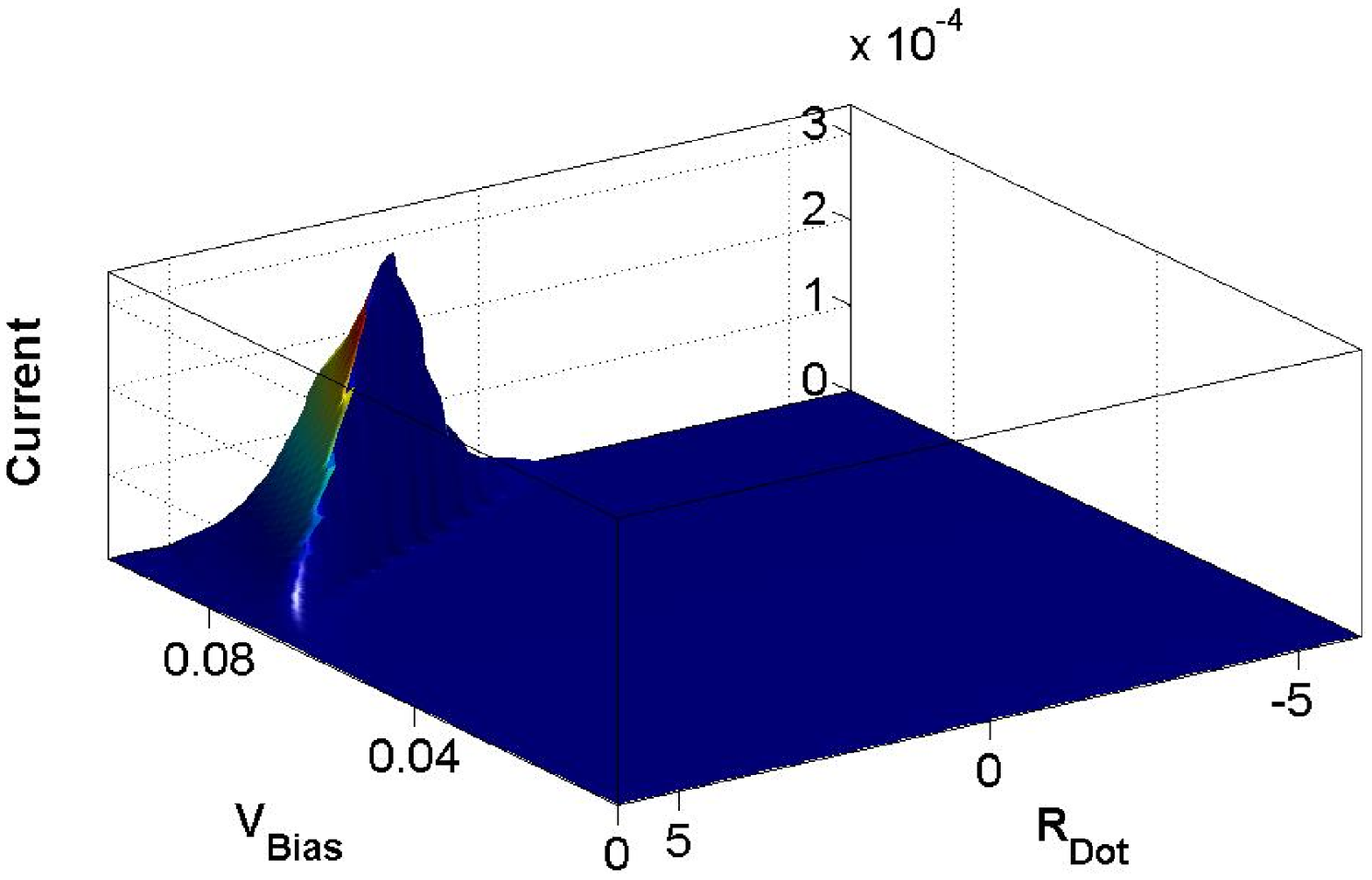}} {%
\includegraphics[width=2.4in,height=2.4in]{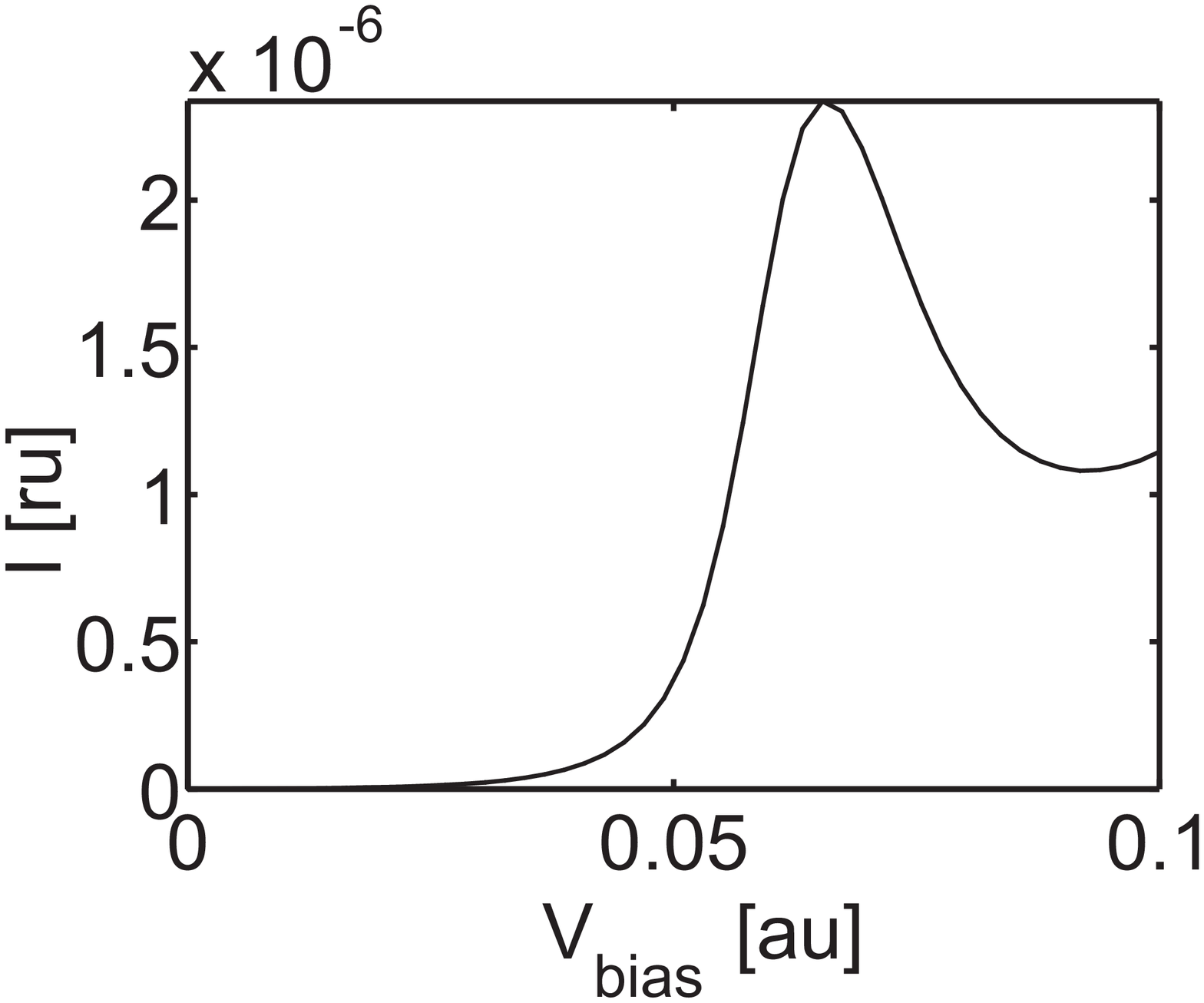}} \sidebyside {} {}
\caption{The current voltage curve of the tunnel transition via
the dipole. Averaging it over the coordinate between the
electrodes is given classically with the Gibbs distribution. The
step increase of current is due to shuttling.} \label{fig11}
\end{figure}
The electrons in the conducting bands of the leads are treated
within the jellium model as shown in Fig. \ref{edm}. The
transmitted electronic waves of the source and drain electrodes
can interfere constructively on the quantum dot, which is located
near the inter-electrode center of the sequential tunnel
junctions.  The Fermi electrons create a maximum density on the
dot, provided the condition of resonance with the dot quantum
level is fulfilled. This effect dominates in the tunnel gap
center, as well as in close proximity to the drain electrode,
which screens a negative charge more effectively thus serving as a
barrier to electron tunneling. When the bias voltage increases,
the resonance level of the dot may shift downward the transparency
window allowed by the Pauli principle, and that may entail a
negative differential conductivity in concert with suppression of
the electron interference, as featured in Figs. \ref{fig10},
\ref{fig11}.

\section{Tunneling Optical Traps}

Recent progress in the atomic microchips industry, has stimulated
great interest in studies of neutral ultracold gases \cite{Lin}.
The ultra cold atomic samples are typically produced in
magneto-optical traps, then loaded into either stable microtraps
or atomic guides, which employ microfabricated structures of
current-carrying wires at surface substrates. These chips offer
great promise for continuously improving the functioning of a
variety of atomic-optic devices, including matter-waves
interferometers, double well potentials for atomic Josephson
junctions, and Fabry-Perot resonators for coupling cavities with
atoms. However, a number of physical mechanisms stipulate
fundamental limitations and hinder immediate applications of these
devices.

The micro-devices require large gradients of magnetic fields for
manipulation of cold atoms. Thus, large currents, up to $1$ $Amp$,
create significant magnetic forces in close proximity $(\sim 1\mu
m)$ to surfaces. Apart from collisional losses in a traped thermal
atomic cloud, strong thermal- and Johnson-noises due to current
fluctuations inevitably induce additional losses, by heating and
pushing atoms to the surfaces. Recently created combined magnetic
and electrostatic traps \cite{Kruger} suffer from similar
shortcomings: Up to $100$ $V$-high bias voltages applied between
$\mu m$-scale distant electrodes are also accompanied by trap
fluctuations that prevent an accurate deposition of the atoms
close to the electrode surface.
\begin{figure}[th]
\caption{The schematics of tunneling optical trap (TOT): the evanescent
field of the blue detuned light repels the atoms from the laser irradiated
substrate and the electrical terminals.} \label{TOTa}\sidebyside
{\includegraphics[width=2.5in]{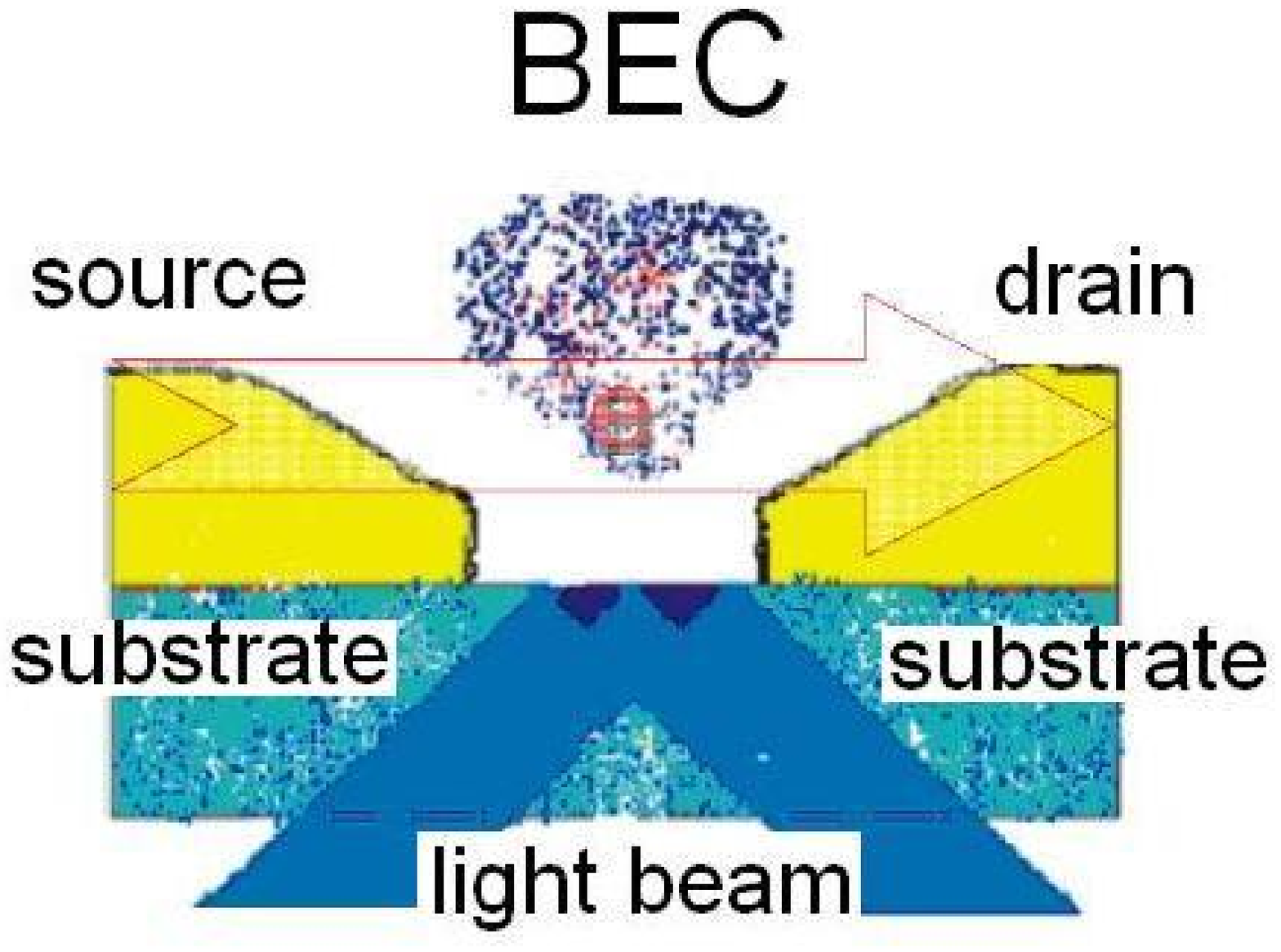}}
{\includegraphics[width=2.5in]{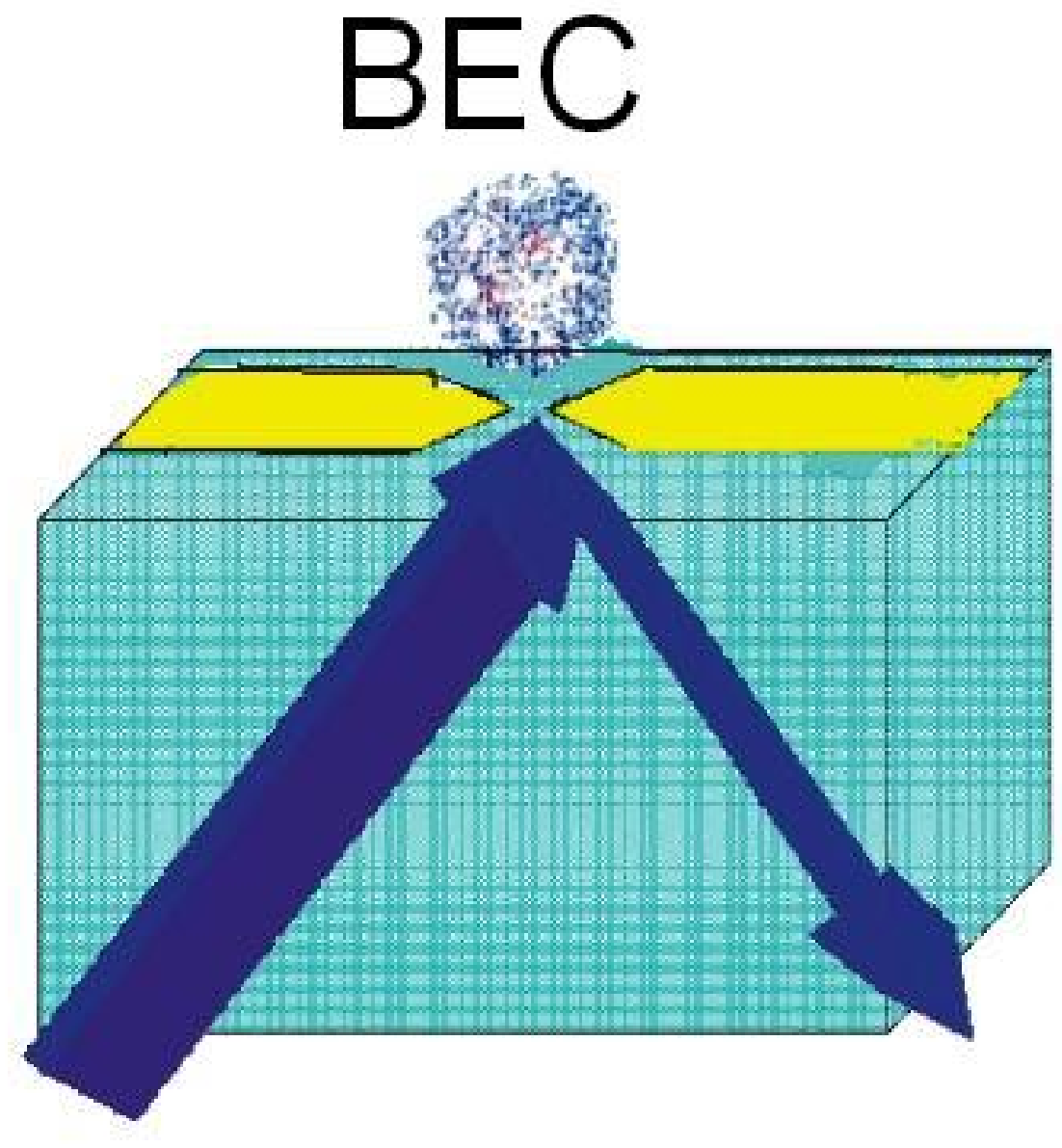}} \sidebyside {}{} \caption{The
current carried by terminals on the surface tend vice versa to attract the
atoms to the inter-electrodes zone of maximum field. The TOT current is
easily controlled by the evanescent light that plays the role of the gate
electrode.}\label{TOTb}
\end{figure}

Theoretically, the electro-optical trapping (TOT), shown in Figs.
\ref{TOTa}, \ref{TOTb}, is a much more appealing scheme for
maintaining and manipulating atoms as compared to the MOT, since
the electric coupling $dE_{s}$ or $e\phi $ in such traps does not
involve the atomic fine structure and is much stronger than the
magnetic coupling. This reduces by many orders of magnitude the
required voltage applied between the electrodes in order to create
a trap of a same depth. Moreover, the TOT is free from the
Majoranna spin-flip losses. These facts could enable one to
eliminate the most of noises\footnote{The smaller the current and
voltage applied, the smaller the Schottky shot noise.} and to
improve the trap control.

Contemporary standards for labs and industry require construction
of TOT in planar geometry based on the evanescent laser fields.
The dipole potential for evanescent waves reads:
\begin{equation}
U=-0.5\alpha _{s}{E_{s}}^{2},  \label{pol}
\end{equation}
where  the sign of the polarizability $\alpha _{s}$ depends on the
sign of the detuning from resonance. At blue detuning, the field
tends to expel the irradiated atoms from the region with high
intensity of the light (see appendix). The advantage of blue
detuning the evanescent field is that the atoms spend most of the
time in the field-free region, and hence they are less affected by
spontaneous radiation and heating. The attraction of atoms to the
tip's neighbourhood is due to the electrostatic source-drain
field. The separation $0.1-10\mu m$ between the source-drain
electrodes allows the elimination of leakage currents and to
facilitate cooling by electron transport through the resonance
states of the TOT atoms.

Up-to-date techniques of cooling, including radio-frequency
evaporation, optimal control of current, degenerative feedback,
and, last but not the least, adiabatic laser cooling \cite{Chen},
have been purposely developed, and now they enable one, in
principle, to maintain the ultra-cold atoms in a TOT. Sisyphus
cooling by the blue-detuned light may provide the required
dissipation in the TOT and further reduce a diffusive heating
produced by the electron current and atomic collisions. With the
repulsive light forces, which push atoms towards the dark regions
near the trap center where the radiation losses should be minimal,
the cold atoms can be significantly compressed adiabatically, thus
yielding a background-free sub-10-nm spot \cite{David}. In
addition, constructive electron interference in these regions
provides a maximum transparency of the resonant tunneling, which
is required in order to minimize losses due to current driven
heating of the atomic cloud.

Moreover, delivering the already cooled atoms close to the surface
does not pose a serious problem. The most routine method of
loading is cold-cloud  transport directly from MOT to TOT. Loading
atoms on the fly by photodesorption from the surfaces of a glass
cell has been demonstrated recently\cite{Atutov}. In order to
minimize radiation heating and collisional losses, the ultra cold
atoms can be isolated in a dark spot near the tunnel gap leads and
the substrate. In fact, a thermal cloud exhibits loss at a
distance larger than the size of a compact condensate, because the
proximity to the surface can provoke cloud
evaporation\cite{Harber}. Then, the interface can be used to
selectively absorb higher energy atoms. Recent MOT experiments
\cite{Lin} demonstrate that by tuning the potential it is possible
to bring ultra-cold atoms to a distance $0.5\mu m$-close to the
surface. This is the case where the TOT can manifest its
excellence.

In addition, the evanescent field of the TOT is concentrated near
the electrode apex, thus providing an additional gain in repulsion
of the cold atoms to the dark spot between the electrode tips.
Therefore, the TOT protects the quantum dots better against
surface losses. Moreover, if the de Broglie wave length of atoms
is larger than the correlation length of the surface roughness,
reflection of the cold remnants occurs elastically. A movable
quantum dot starts to oscillate, being trapped between the voltage
biased tips. The adiabatic dynamics, in which the repulsion of the
evanescent laser field compensates the electrostatic attraction to
the tips, mimics Franklin's Bell oscillations with the shuttling
mechanism of conductivity through the charge BEC.

The strength of attraction between the dot and the source-drain
electrodes depends on the ground state polarizability,
$\alpha_{s}$. For example, the $\alpha _{s}$ is about
$80[{mHz}/{V^{2}}/{cm^{2}}]$ for $Rb^{87}$. For a $\mu m$-scale
inter-electrode gap,  $D\sim 1\mu m$, the voltage bias $V\sim
30mV$ produces an external field pulling the atoms into a
$U_{trap}=\alpha _{s}(V/D)^{2}\sim 8$ KHz-depth trap. For an atom
de Broglie length $\lambda _{dB}\sim \hbar /\sqrt{2MU_{trap}}\sim
0.1\mu $ and a trap size of the order of $D$, we expect that
approximately $N=(D/\lambda _{dB})^{3}\sim 10^{3}$ atoms can be
put into the quantum degenerate regime, provided their protection
against sticking and colliding with surfaces is efficient. The
repulsion potential of the atoms is due to the blue-detuned
evanescent field. The corresponding dipole force compensates for
the electrostatic attraction of the charged ultra-cold particle to
the leads and to the substrate.

The density of the atoms, which should approach the leads as close
as possible without sticking, significantly influences the TOT
resistance $R$. Therefore, the repulsion of atoms has to be
controlled by the attenuation of the evanescent laser field.
Typically the laser beam of $1-100mW$ power can be focused on the
interface to allow ultra-cold atoms to levitate above it.

Levitation of $Cs$ atoms in the evanescent field has been
demonstrated in \cite{Grimm}, where the the optical dipole
potential created by $1W$-laser has been utilized for trapping the
atoms far from the surface. The exponential profile of the
potential decreases at a half-wave length $\lambda _{blue}/2\sim
250nm$. At these distances the optical field compensates for the
long-range attraction induced by the electrostatic polarization
and the Casimir-Polder potential
\begin{equation}
U_{Casimir-Polder}\simeq \frac{c_{4}}{R^{4}},\hskip0.1inc_{4}\sim 1\mathop{}
[{nK}\mathop{}{\mu m^4}].  \label{KP}
\end{equation}
which exists owing to spontaneous electromagnetic field
fluctuations for neutral atoms and ensures the existence of a
stationary point in the net potential field. The Casimir-Polder
interaction for a $Rb^{87}$ atom located at a distance $\sim
0.5\mu m$ from the surface is equal to the polarization potential
Eq. \ref{pol} induced by $15mV$ voltage of the source-drain
terminals. The surface repulsion due to the evanescent field is of
the same order of magnitude for other alkali atoms $K,Na,Rb$,
provided the corresponding matter constants and the optical wave
lengths are taken into account. For the reader's convenience, in
an Appendix we quote the formula for the dipole potentials in the
laser field.

\begin{figure}[tbp]
{\includegraphics[width=3 in]{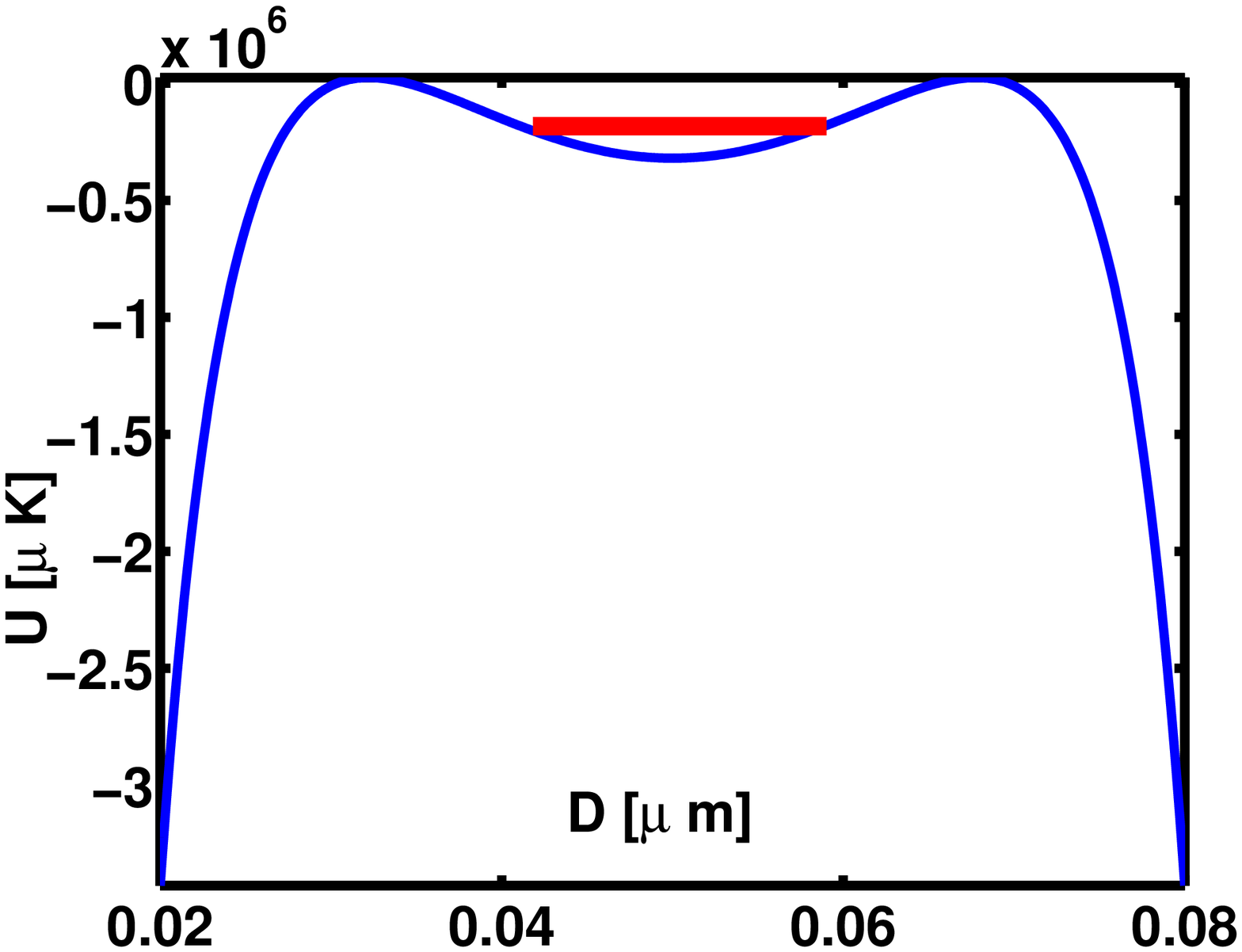}}
\narrowcaption{Schematics of the electro-optical potential
including the Casimir-Polder attraction of eq. \ref{KP} for
$Rb^{87}$ atom between leads. The evanescent field, that decreases
exponentially on its half wave-length expels atoms away from the
surface to the location where attractive forces balance the light
pressure. The line in the well cartoons the trap occupation.}
\end{figure}

Both the optical and the electrical trapping have already been
demonstrated separately. The double evanescent wave trap for atoms
has been proposed in \cite{Ovch} and has been demonstrated in
\cite{Grimm} for Cs atoms. Bezryadin and coworkers\cite{Bezr} have
reported a method of fabrication for nanoscale controllable break
junctions, in which the polarized nano-clusters, such as Pd
clusters, have been trapped between the electrodes. The
nano-clusters were self-assembled in the region of the maximum
field in order to produce a wire connecting the tips \cite{Bezr}.

Both types of trapping, by optical and by electrical potentials,
eventually can be combined into a common protocol of
cross-coupling ultra-cold atoms and electrical circuits. Such a
cross coupling of the electron current and the ultra-cold atoms in
matter wave-guides is proposed here using the example of hollow
optical fibre guiding. In this scheme, a laser field, detuned
either to the blue or to the red wing of the atomic transition,
allows the atoms to be guided through the fibre capillary. The
technique has already been routinely used \cite{Dall}.

Here we suggest that experimentalists address the ultra-cold
metallic gas cloud in the capillary channel
$\mathit{electrically}$ in order to measure a tunneling I-V curve.
To this end, the voltage bias could be led to the cold cloud
through a lateral wire, which crosses the bulk of the
optical-guide wall and immediately breaks-off inside the
capillary. The blue-detuned evanescent field repels the ultra-cold
atoms towards the capillary axis. The ponderomotive field
potential accelerates the tunneling electrons and facilitates
charging of the trapped atoms. This scheme features an effective
reduction of the collisional loss and protection from leakage of
currents. The optical field protects the electrical reservoirs of
the source-drain leads from any disturbances by tunnel electrons.
The last but not least advantage is that the ultra-cold atoms are
trapped inside the capillary in the dark spot, which is affected
neither by heating caused by the spontaneous decay nor by induced
radiation processes.

The TOT scheme relying on the wave-guides also has a practical
advantage over the planar design, since it requires a smaller
number of microfabricated components. The heart of the setup is
the quantum dot, which is created by self-assembly of the
ultra-cold cloud in a dark spot between the electrodes. The
self-assembly due to the optical dipole force and the electrical
polarization potential is very useful for fabrication of TOT
devices at such a small length scale. This scheme of
cross-coupling merits a detailed consideration, since it should
allow one to discover the true electrical resistance of atomic
BEC.\footnote{Prof. V.L. Ginzburg once said that any synthesized
atomic or molecular combination is a valid candidate in .... the
quest for a superconductive state.} Theoretically, the BEC
collective state could be described by a composite order
parameter, which is formed by the diatomic molecules and
superconducting BCS pairs. It would be advantageous to map the
atomic wave coherences to the coherent electric responses (echoes)
of the normal or superconducting leads by applying the voltage
bias to the quantum degenerate ultra-cold gas in the TOT.

We conclude this section by listing the evident TOT advantages:

\begin{itemize}
\item Small density of states, typical of the TOT atoms, reduces the leakage
currents, the dissipative tunneling, and noises in measurements.

\item High protection of atoms in the TOT from the influence of the environment
favors coherence of their electrical response (nutations and echoes) to
pulsed bias.

\item Controllable double-, triple-, or, in principle, multiple-well
potential as well as current oscillations in the atomic Josephson
transitions can be achieved.

\item Precise control of matter wave interferometry is foreseen.

\item Combined addressing of the quantum states of the TOT both optically and
electrically is possible.
\end{itemize}

In brief, we propose the TOT electro-optical setup in which
electrical measurements have a high quality factor combined with
the coherence of an all optical experiment. The TOT can be easily
incorporated in electrical circuits as a nonlinear element
ensuring a scalability of the architecture. The quantum dot
isolation in the TOT will protect entanglement of quantum states
thus permitting field programmable gates arrays.

\section{Coherence of electron transport via double wells}


Thus far we have made no reference to a wave-like motion of the
NEM-SET devices. However, the coherent behaviour of the trapped
quantum dots will become increasingly important with the NEM-SET
made smaller and colder. The wave function $\Psi $ formalism is
useful as a description of quantum dots only when their de Broglie
wavelength $L_{d}$ is comparable with the Fermi length of
electrons, that is $L_{d}\gamma \sim 1$, and we can rely on the
tunnel curves. At the temperature when quantum degeneracy occurs,
the self consistent charge Eq.\ref{char} transported through the
SET junctions via the shuttle mechanism is not smeared over the
tunnel terms. As demonstrated in Sec. 7 and 8, a peak of the
charge distribution is centered between the electrodes and the
Coulomb blockade abruptly breaks this symmetry of the tunnel term
when the bias voltage $V\geq U_{c}$. The single-well bond then
bifurcates into a double-well potential, because the barrier
between the wells brings a gain in the Coulomb energy. The
corresponding self-consistent charge is imposed by the condition
of detailed balance between the electron flows and an irreversible
coupling $\Gamma _{\sigma }$.

Tunneling of non-correlated itinerant electrons through the lead
interfaces at the rate $\Gamma _{\sigma }$  provides irreversible
dissipation, which is required in our approach\footnote{as well as
in the orthodox theory of Coulomb blockade \cite{Kulik},
\cite{Aver}} by the detailed balance condition. The electron flows
charge the resonant state of the dot (see Eq. \ref{QF}) thus
forming the tunnel curves $U_{\sigma }$ (where $\sigma =l,r$),
provided the charge is located in the center between the
electrodes. The  electron scattering and tunneling are the
dissipating mechanisms that ensure stability of the adiabatic
dots' dynamics. The dots can also relax their kinetic energy by
collisional losses, scattering, spontaneous emission, or laser
cooling. In quantum gases, binary scattering of atoms usually
results in a shift of the energy level, which is proportional to
the product of $s$-scattering length $a$ and the gas density
$\rho$. The interplay of the quantum spreading and scattering of
the dots sitting in a single-well potential $U(x)$ is accounted
for by the Gross-Pitaevskii equation for the "condensate" wave
function $\Psi $ \cite{Pita} as
\begin{equation}
i\hbar \dot{\Psi}=\frac{{p^{2}}}{{2\tilde{M}}}\Psi +U(x)\Psi +\frac{{4\pi
\hbar ^{2}a}}{\tilde{M}}\left\vert \Psi \right\vert ^{2}\Psi ,  \label{GP}
\end{equation}
where the $\hat{p}=i\hbar \frac{{\partial }}{{\partial x}}$ is the
momentum operator, the $\tilde{M}$ is the effective mass of a dot.
The scattering length $a$ has to be smaller than the average
distance between dots in the trap, this serves as a criterion of
validity of the Gross-Pitaevskii equation. The quantum dispersion
of the dot $\frac{1}{2\tilde{M}}\hat{p}^{2}$ brings about the
major spreading mechanism in the degenerate quantum regime. The
electron transport critically depends on the bias voltage $V$: For
$V<U_{c}$ the adiabatic potential $U$ features a single well,
which for $V>U_{c}$ bifurcates into a double-well potential with
degenerate states located in the well minima.

Should one disregard the $s$-scattering off a small density of
states $\left\vert \Psi \right\vert ^{2}$ in a dilute quantum gas,
the resulting linear Schr\"{o}diger equation would describe the
dynamics of the $\Psi$-state in the potential $U$. The Landau
bifurcation then emerges from casting the potential $U(x)$ of Eq.
\ref{AP0} into a Taylor series over a small deviation $x$ from the
bifurcation point:
\[
U(x)=\tilde{M}(\Omega ^{2}-\Sigma ^{2})x^{2}+\Xi x^{4}.
\]
Parameters $\Sigma $ and $\Xi $ are defined by the condition of
Coulomb blockade. They depend on the temperature, the bias
voltage, and the coupling to electron reservoirs, that all
influence the criticality of quantum dynamics. The quantum
degenerate regime is meaningful when the dot temperature is
smaller than the Coulomb energy. The heat supplied to the dots by
the bias voltage has to be dissipated in order to reach the
degeneracy point. The required relaxation is provided by the
radiation channels via the Josephson plasmons of the dot tunneling
across the Coulomb barrier. The tunnel transitions split the
degenerate energy levels corresponding to the dot motion in the
wells' minima of the bifurcated potential. The frequency of
splitting is just the Josephson frequency quantum.

Alternatively the quantum dynamics on the tunneling curve can be
invoked by the wave functions $\Psi _{l}$ and $\Psi _{r}$ of the
dots, each of which resides in its own well. The functions
$\Psi_{l}$ and $\Psi _{r}$ obey the system of equations
\begin{eqnarray}
i\hbar \dot{\Psi _{l}} &=&\hat{H}_{l}\Psi _{l}+\Delta _{l}\Psi_{r}
\mathletter{a}, \\
i\hbar \dot{\Psi _{r}} &=&\hat{H}_{r}\Psi _{r}+\Delta _{r}\Psi
_{l} \mathletter{b},  \label{Joseph}
\end{eqnarray}%
where $\sigma =l,r$, and the quantities $g_{\sigma }=4\pi \hbar^{2}a_{\sigma
}/\tilde{M}$ are related to the corresponding $s$-scattering lengths of the
matter waves with Hamiltonians $\hat{H}_{\sigma
}=\frac{1}{2\tilde{M}}\hat{p}^{2} +U_{\sigma}+g_{\sigma }\left\vert \Psi
_{\sigma }\right\vert^{2}$. Equations $59a,b$ can be considered as a
generalization of the Gross - Pitaevskii equation of the interacting charged
Bose gas. Alternatively, it can be obtained from the many-body Hamiltonian
of the field theory. The mean field Hartree dynamics is presented in Eqs.
$59a,b$. The tunnel terms $U_{l,r}$ are modified by the collisional shifts
due to $s$-scattering of particles in the same well. The $\Delta _{l,r}$
defined below implies the inter-wells' scattering. For the charged Bose gas
the interaction Hamiltonian of the field theory reads
\begin{equation}
\hat{H}_{{\mathop{\rm int}}}=\int \int \hat{\Psi}^{\dag }(x)\hat{\Psi}^{\dag
}(x_{1})\varphi (x,x_{1})\hat{\Psi}(x_{1})\hat{\Psi}(x),  \label{Hamint}
\end{equation}
where $\varphi (x,x_{1})$ denotes the Green's function of the Poisson
equation for the electrostatic potential in the inter-electrodes zone. We
represent the field operator of the Bose field as
\[
\hat{\Psi}(x)=\Psi _{l}(x)+\Psi _{r}(x)+fluctuating\mathop{}fields,
\]
where the average of the fluctuating part tends to zero in the
thermodynamic limit. The overlap integral $\Delta $ measures the
rate of particle tunneling through the barrier. The integral
$\Delta $ is obtained from the variational derivative of the
interaction Hamiltonian Eq. \ref{Hamint}
\[
\frac{\delta \hat{H}_{\mathop{\rm int}}}{\delta \hat{\Psi}^{\dag
}(x)}=\int {dx_{1}\varphi (x,x_{1})\left\langle {\hat{\Psi}^{\dag
}(x_{1})\hat{\Psi} (x_{1})}\right\rangle }.
\]
The dependence of  $\varphi$ on $x$ is slow and we can disregard
it as compared to the function of the difference $x-x_{1}$
modelled by the Dirac $\delta $-function, so we have
\begin{equation}
\Delta _{l,r}\approx \int {dx_{1}\varphi (x_{l,r},x_{1})\Psi _{r}^{\ast
}(x_{1})\Psi _{l}(x_{1})\approx \frac{{4\pi \hbar ^{2}a}}{\tilde{M}}\Psi
_{r}^{\ast }(x_{l,r})\Psi _{l}(x_{l,r}).}  \label{Delta}
\end{equation}
The two states $\Psi _{l}$ and $\Psi _{r}$ are macroscopically
distinct if their separation is of the order of the wavelength.
For instance, ultra-cold rubidium $Rb$ atoms oscillating in a TOT
with frequency $1kHz$ have a wavelength of about $1\mu m$. The
coherence length limits the maximum barrier width that allows
atoms to be hybridized between the wells. The model describes
screening in the charged Bose gas with Josephson plasmons. The
current through the degenerate quantum gas reads
\begin{equation}
J_{l}=\frac{2e}{h}\left\{ \Psi _{l}^{\ast }\frac{d\Psi
_{l}}{dx}-\Psi _{l} \frac{d\Psi _{l}^{\ast }}{dx}\right\} .
\end{equation}
The Josephson current is obtained by multiplying Eq. \ref{Joseph}
by the conjugated wave function $\Psi _{l}^{\ast }$ and
integrating by parts. This yields the flow of the charged Bose gas
with the current
\begin{equation}
J_{l}=\frac{{2e}}{h}\Delta {\mathop{\rm Im}\nolimits}\left\{ {\int
{dx} \mathop {\mathop {\Psi _l}\limits^*
(x,t)}\limits^{{}}\Psi_{r}(x,t)} \right\} .
\end{equation}
The overlap of the wave functions oscillates at a frequency
$\Delta =\Delta_{l}\approx \Delta _{r}$. For weak superconductors,
the tunneling frequency $\Delta $ is controlled by the source
drain voltage. The alternating current (ac) is zero below the
threshold voltage and it oscillates after the threshold with an
amplitude growing with the bias voltage. Existence of ac at the
frequency $\Delta $ immediately indicates the broken symmetry of
the bond potential. The frequency of the current is proportional
to the source-drain voltage and, thus, the SET radiates an
electromagnetic field. This radiation causes relaxation that
ensures a back-action mechanism for establishing an equilibrium in
the composite system.

It is worthwhile to note that the charged Bose gas trapped in the double
well potential $U_{\sigma }$ of Eqs. $59$ behaves as an inverted Josephson
junction (N-S-S-N). The super-current, which accompanies the matter wave
coherence, is induced  between the degenerate resonance states of the
adjacent wells at the frequency of the tunnel splitting $\Delta \ll \Omega$.
While the shuttling frequency $\Omega$ cannot be displayed directly because
of a large response time, as is typical of tunnel junctions (whose frequency
cutoff is much smaller than the vibrational frequency even for nano
junctions). The coherent oscillations of the Josephson current can be
observed by virtue of their slow frequency $\Delta \sim V $ which is
robustly controlled by the bias voltage.

\section{Summary}

The step-wise and negative differential resistance regions of the
current-voltage curves observed in the molecular $C_{60}$
transistor \cite{Park} are explained by the field effect, in which
the voltage bias of the source-drain leads intensifies the NEM
effective temperature. With increasing bias, the field splitting
and the inhomogeneous broadening modify the transparency spectra
of electron scattering. We have obtained a good qualitative
agreement between our simple adiabatic models and the differential
conductance observed experimentally. Our formula also displays the
crossover between the internal mode ($\sim 33$ meV) and the
bouncing-ball mode ($\Omega \approx 5$ meV) of the differential
conductivity in the molecular SET.  With further increase of the
electrostatic interaction, the shuttling mechanism of conductivity
replaces the tunneling regime. Then, the Coulomb energy growth
destroys the bonding symmetry of the single well potential. The
broken symmetry of the molecular vibrations in back-action cures
the shuttling instability. The shuttle regime is characterized by
shot noise reduction and by coherence of the oscillating current.
The primary quantization picture demonstrates that the current
oscillates at the frequency of the ground state splitting $\Delta$
and the amplitude of this oscillation grows proportionally to
$\Delta $.

We have developed an unified adiabatic approach allowing one to
tackle transport problems in traps of different geometry. The
magnetic and electrical fields, charge screening, and other
factors (a spin-orbit interaction, hyperfine structure,
\textit{etc}) can influence the quantum dot paths within an easily
tractable  Breit-Wigner-resonance approximation for the electron
scattering.  The utility and universality of the tunnel terms
concept are confirmed for the phenomenological and 'ab-inition'
theories of the shuttling instability.

The shuttling in a quantum gas is relevant to electron transport
in the presence of relaxation and the Coulomb blockade. The
Coulomb blockade entails broken symmetry of molecular potentials
$U_{\sigma }$ ($\sigma =l,r$) when a single well bifurcates into a
double well. Tunneling of the charged Bose gas in a double well
between biased electrodes creates a current which is subject to
Josephson oscillations. This ac generates an electro-magnetic
field and thus providing an additional mechanism of dissipation.
Thence, the broken symmetry and coherent oscillation due to
molecular vibrations ensure the necessary and sufficient
conditions legitimizing the present scenario of bond bifurcation.

The adiabatic theory of electron transport in the Breit-Wigner
approximation may be of more than academic interest. It can help
one to devise the TOT protection of the NEM - SET systems against
decoherence. The TOT technology is better suited for molecular
optoelectronics due to a low noise in combination with protection
control. The connection between the TOT conductivity and quantum
Franklin's Bell paradigm is discussed. The TOT design for avoiding
the dissipative " roadblocks", could serve as a road map toward a
new generation of optical SET, that should enable electrical
non-demolition measurements on the quantum threshold. It would be
of great interest to measure the resistance of a TOT comprising of
BEC molecules and BCS atomic pairs.


\begin{acknowledgments}

With the present paper we pay a tribute to our friend and teacher
A.P.Kazantsev. He made the seminal contribution to the now
enormous field of activity of mechanical action of light on
neutral atoms. He often prophesied the future development for
years ahead. Four decades ago he recognized the significance of
radiation from accelerated charged particles near metallic
surfaces \cite{KS} that appears to be of importance for
dissipative mechanisms in TOT electron transport.

The authors are indebted to colleagues for numerous discussions.
In particular we thank A.Bezryadin for his preprint \cite{Bezr}
which he sent to us, to I. Novobrantsev for keen interest and
encouragement, and G.Surdutovich for his valuable remarks and
references. We are thankful to RFBR and the program of scientific
schools for the financial support.  The hospitality of INF in
Ferrara (Italy) is greatly acknowledged by one of the authors
(A.R.), especially,  to S. Atutov and  R. Calabrese for a kind
invitation. The collaboration with the experimental team at the
winters 2000-2001 has led to the ideas developed therein.
\end{acknowledgments}

\begin{glossary}
\term{AOC} Anderson orthogonality catastrophe \cite{AOC}. Zero overlap between ground
states of surface vibrations: e.g. the original state and charge induced
configurations, where phonons are shifted by the tunneling electron.

\term{BEC} Bose-Einstein condensate. Degenerate state of an
ultra-cold ensemble,
 i.e. a coherent, single-mode, bright atomic source of zero momentum $p=0$.

\term{MOT} Magneto Optical Trap. The well established technique
for Doppler cooling and trapping of a thermal cloud of cold atoms.

\term{NDR} Negative differential resistance.

\term{NEMs} Nanoelectromechanical systems. Composite mesoscopic and nanoscale devices designed
for a new functionality.

\term{SET} Single-electron transistor. Double tunnel junctions with a central island
serving as a gate electrode.

\term{SSET} Superconducting Single-electron transistor.  The same
as SET, but the bulk of the electrodes are superconducting.

\term{TOT} Tunneling Optical Trap.


\end{glossary}
\chapappendix{Radiation pressure}

Demonstration of levitation of micron-sized latex particles by
radiation pressure dates back to 1970 in the experiments reported
by Ashkin \cite{Ashkin}. The average force accelerating (or
slowing down) atoms in a laser field was derived by A.Kazantsev in
1972 \cite{Kaz}. Later in 1972-1974 he classified the optical
forces as spontaneous, induced and mixed. In particular, it was he
who first presented the dipole potentials for velocity broadened
lines of resonance atoms in the logarithmic form

\begin{equation}
U=\frac{\delta}{2}\ln\left[ {1+\frac{G^2}{\delta^2+\gamma^2}} \right]
\end{equation}

here $\delta=\Omega-\Omega_0$ is the detuning from resonance,
$\gamma $ is the atomic linewidth,
$\Omega,\hskip0.1in\Omega_0,\hskip0.1in G=dE$ is the frequency of
light, the resonance transition and the Rabi frequency
respectively, $d$ is the dipole moment of the transition (in units
$\hbar=c=1$), and $E$ is the laser field amplitude. At large
detunning, the dipole potential takes the canonical form
(Askaryan, 1962)
\[
U=-0.5\alpha E^2,
\]
where the optical polarization is
\[
\alpha=-\frac{d^2}{\delta}.
\]
This potential repels the atom from antinodes of the blue detuned
field $\Omega \geq \Omega_0$. The potential of red detuned light,
$\Omega \leq\Omega_0$, vice versa attracts the dots to the field
antinodes. Use of both methods allows  guiding and trapping of
atomic matter waves.

\end{document}